\documentclass[twocolumn,showpacs,amsmath,amssymb,aps,prx,footinbib,floatfix,superscriptaddress,longbibliography]{revtex4-1}
\usepackage{graphicx}
\usepackage{braket}
\usepackage{savesym}
\usepackage{amsfonts}
\usepackage{bm}
\usepackage{color}
\usepackage{subfigure}
\usepackage{wasysym}
\usepackage{upgreek}
\usepackage{float}
\usepackage{dsfont}
\usepackage{mathtools}
\usepackage{multirow}
\usepackage[usenames,dvipsnames,svgnames,table]{xcolor}
\usepackage{hyperref}
\usepackage{hhline} 
\usepackage{multirow} 
\usepackage[normalem]{ulem}

\usepackage{makecell}

\usepackage[english]{babel}

\hypersetup{
	colorlinks=true,       
	linkcolor=VioletRed,  
    citecolor=JungleGreen,        
    filecolor=Orchid,      
    urlcolor=black           
}
\begin{document}

\title{Variational Neural Annealing} 

\author{Mohamed Hibat-Allah}
\email{mohamed.hibat.allah@uwaterloo.ca}
\affiliation{Vector Institute, MaRS  Centre,  Toronto,  Ontario,  M5G  1M1,  Canada}
\affiliation{Department of Physics and Astronomy, University of Waterloo, Ontario, N2L 3G1, Canada}

\author{Estelle M. Inack}
\affiliation{Perimeter Institute for Theoretical Physics, Waterloo, ON N2L 2Y5, Canada}
\affiliation{Vector Institute, MaRS  Centre,  Toronto,  Ontario,  M5G  1M1,  Canada}

\author{Roeland Wiersema}
\affiliation{Vector Institute, MaRS  Centre,  Toronto,  Ontario,  M5G  1M1,  Canada}
\affiliation{Department of Physics and Astronomy, University of Waterloo, Ontario, N2L 3G1, Canada}

\author{Roger G. Melko}
\affiliation{Department of Physics and Astronomy, University of Waterloo, Ontario, N2L 3G1, Canada}
\affiliation{Perimeter Institute for Theoretical Physics, Waterloo, ON N2L 2Y5, Canada}

\author{Juan Carrasquilla}
\affiliation{Vector Institute, MaRS  Centre,  Toronto,  Ontario,  M5G  1M1,  Canada}
\affiliation{Department of Physics and Astronomy, University of Waterloo, Ontario, N2L 3G1, Canada}

\date{\today}

\begin{abstract}

Many important challenges in science and technology can be cast as optimization problems. When viewed in a statistical physics framework, these can be tackled by simulated annealing, where a gradual cooling procedure helps search for groundstate solutions of a target Hamiltonian. While powerful, simulated annealing is known to have prohibitively slow sampling dynamics when the optimization landscape is rough or glassy. Here we show that by generalizing the target distribution with a parameterized model, an analogous annealing framework based on the variational principle can be used to search for groundstate solutions. Modern autoregressive models such as recurrent neural networks provide ideal parameterizations since they can be exactly sampled without slow dynamics even when the model encodes a rough landscape. We implement this procedure in the classical and quantum settings on several prototypical spin glass Hamiltonians, and find that it significantly outperforms traditional simulated annealing in the asymptotic limit, illustrating the potential power of this yet unexplored route to optimization.

\end{abstract}

\maketitle

 \section{Introduction}
A wide array of complex combinatorial optimization problems can be reformulated as finding the lowest energy configuration of an Ising Hamiltonian of the form~\cite{lucas2014ising}:
\begin{equation}
     H_\text{target} = -\sum_{i < j} J_{ij} \sigma_i \sigma_j - \sum_{i=1}^{N} h_{i} \sigma_i,
\label{eq:IsingGlassHamiltonian}     
\end{equation}
where $\sigma_i=\pm 1$ are spin variables defined on the $N$ nodes of a graph. The topology of the graph together with the couplings $J_{ij}$ and fields $h_{i}$ uniquely encode the optimization problem, and its solutions correspond to spin configurations $\{\sigma_i\}$ that minimize $H_\text{target}$. While the lowest energy states of certain families of Ising Hamiltonians can be found with modest computational resources, most of these problems are hard to solve and belong to the non-deterministic polynomial time (NP)-hard complexity class~\cite{Barahona_1982}.    

Various heuristics have been used over the years to find approximate solutions to these NP-hard problems. A notable example is simulated annealing (SA)~\cite{Kirkpatrick671}, which mirrors the analogous annealing process in materials science and metallurgy where a crystalline solid is heated and then slowly cooled down to its lowest energy and most structurally stable crystal arrangement. In addition to providing a fundamental connection between the thermodynamic behavior of real physical systems and complex optimization problems, simulated annealing has enabled scientific and technological advances with far-reaching implications in areas as diverse as operations research~\cite{KOULAMAS199441}, artificial intelligence~\cite{hajek1985}, biology~\cite{svergun1999}, graph theory~\cite{johnson1991}, power systems~\cite{abido2000}, quantum control~\cite{PhysRevB.91.201404}, circuit design~\cite{gielen1989} among many others~\cite{hajek1985}.
The paradigm of annealing has been so successful that it has inspired 
intense research into its quantum extension, which requires quantum hardware 
to anneal the tunneling amplitude, and can be simulated in an analogous way to SA
\cite{Santoro_2002,Brooke1999}.

\begin{figure}
    \centering
    \includegraphics[width = 0.8\linewidth]{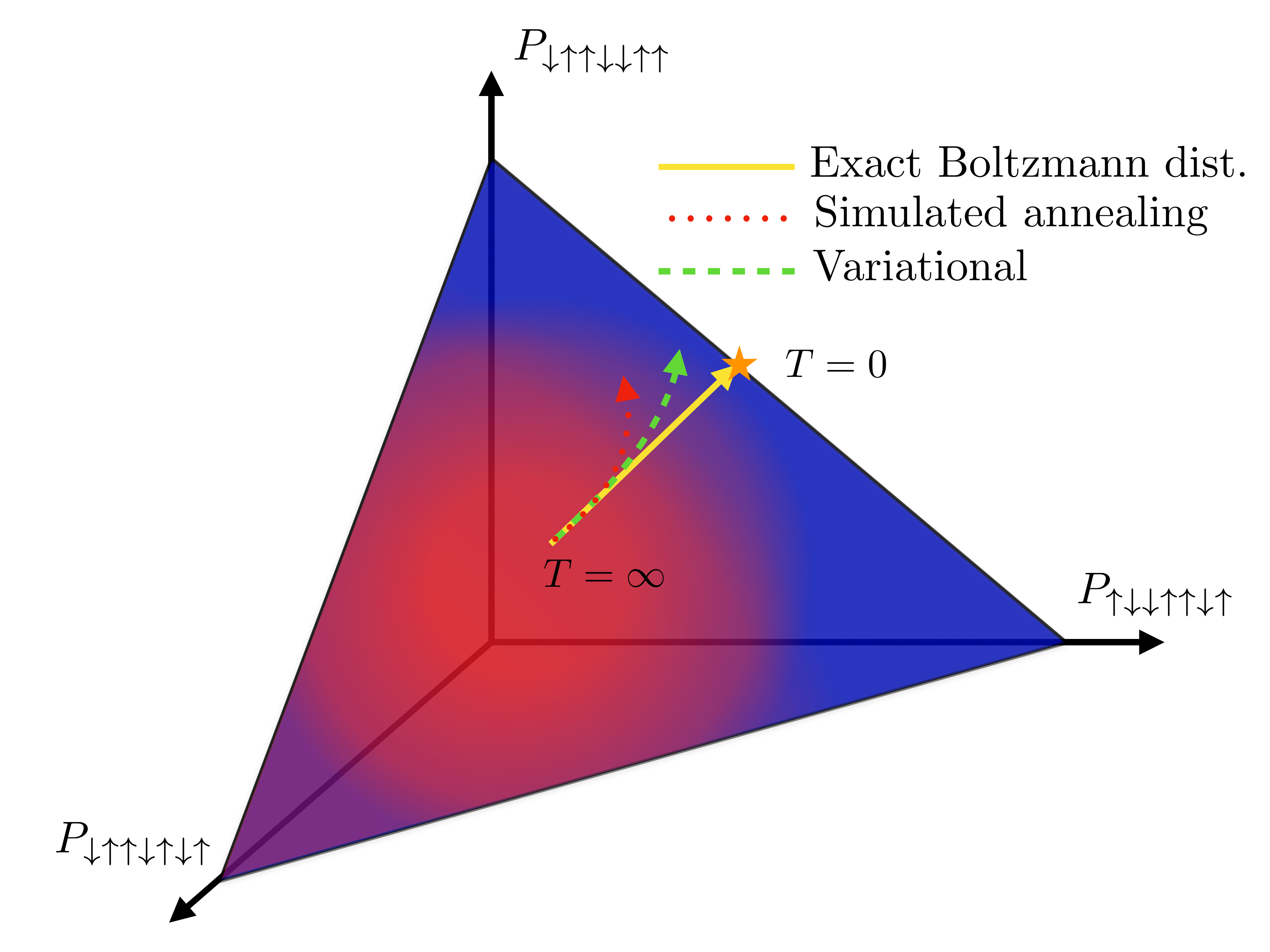}
    \caption{Schematic illustration of the space of probability distributions visited during simulated annealing. An arbitrarily slow SA visits a series of Boltzmann distributions starting at the high temperature (e.g. $T=\infty$) and ending in the $T=0$ Boltzmann distribution (continuous yellow line), where a perfect solution to an optimization problem is reached. These solutions are found either at the edge or a corner (for non-degenerate problems) of the standard probabilistic simplex (colored triangle plane). A practical, finite-time SA trajectory (red dotted line), as well as a variational classical annealing trajectory (green dashed line), deviate from the trajectory of exact Boltzmann distributions. }
    \label{fig:schematicclassicalannealing}
\end{figure}

The SA algorithm explores an optimization problem's energy landscape via a gradual decrease in thermal fluctuations generated by the Metropolis-Hastings algorithm. The procedure stops when all thermal kinetics are removed from the system, at which point the solution to the optimization problem is expected to be found. While an exact solution to the optimization problem is always attained if the decrease in temperature is arbitrarily slow, a practical implementation of the algorithm must necessarily run on a finite time scale~\cite{mitra1986}. As a consequence, the annealing algorithm samples a series of effective, quasi-equilibrium distributions close but not exactly equal to the stationary Boltzmann distributions targeted during the annealing~\cite{delahaye2019} 
(see Fig.~\ref{fig:schematicclassicalannealing} for a schematic illustration). This naturally leads to approximate solutions to the optimization problem, whose quality generally depends on the interplay between the problem complexity and the rate at which the temperature is decreased. 

In this paper, we offer an alternative route to solving optimization problems of the form of Eq.~\eqref{eq:IsingGlassHamiltonian}, called {\it variational neural annealing}. 
Here, the conventional simulated annealing formulation is substituted with the annealing of a parameterized model.
Namely, instead of annealing and approximately sampling the exact Boltzmann distribution, this approach anneals 
a quasi-equilibrium model, which must be sufficiently expressive and capable of tractable sampling.
Fortunately, suitable models have recently been provided by machine learning technology ~\cite{10.5555/3104482.3104610,pmlr-v15-larochelle11a,vaswani2017attention}. In particular, {\it neural autoregressive} models combined with variational principles have been shown to accurately describe the equilibrium properties of classical and quantum systems~\cite{Wu_2019,Sharir_2020,RNNWF_2020,roth2020iterative}.
Here, we implement variational neural annealing using autoregressive recurrent neural networks, and show that they offer a powerful alternative to conventional SA and its analogous quantum extension, i.e., simulated quantum annealing (SQA)~\cite{Santoro_2002}. This powerful and unexplored route to optimization is schematically illustrated in Fig.~\ref{fig:schematicclassicalannealing}, where a variational neural annealing trajectory (dashed green arrow) is shown to provide a more accurate approximation to the ideal trajectory (continuous yellow line) than a conventional SA run (dotted red line).


\section{Variational classical and quantum annealing}
\label{sec:VCA}

\begin{figure*}
    \centering
    \includegraphics[width = \linewidth]{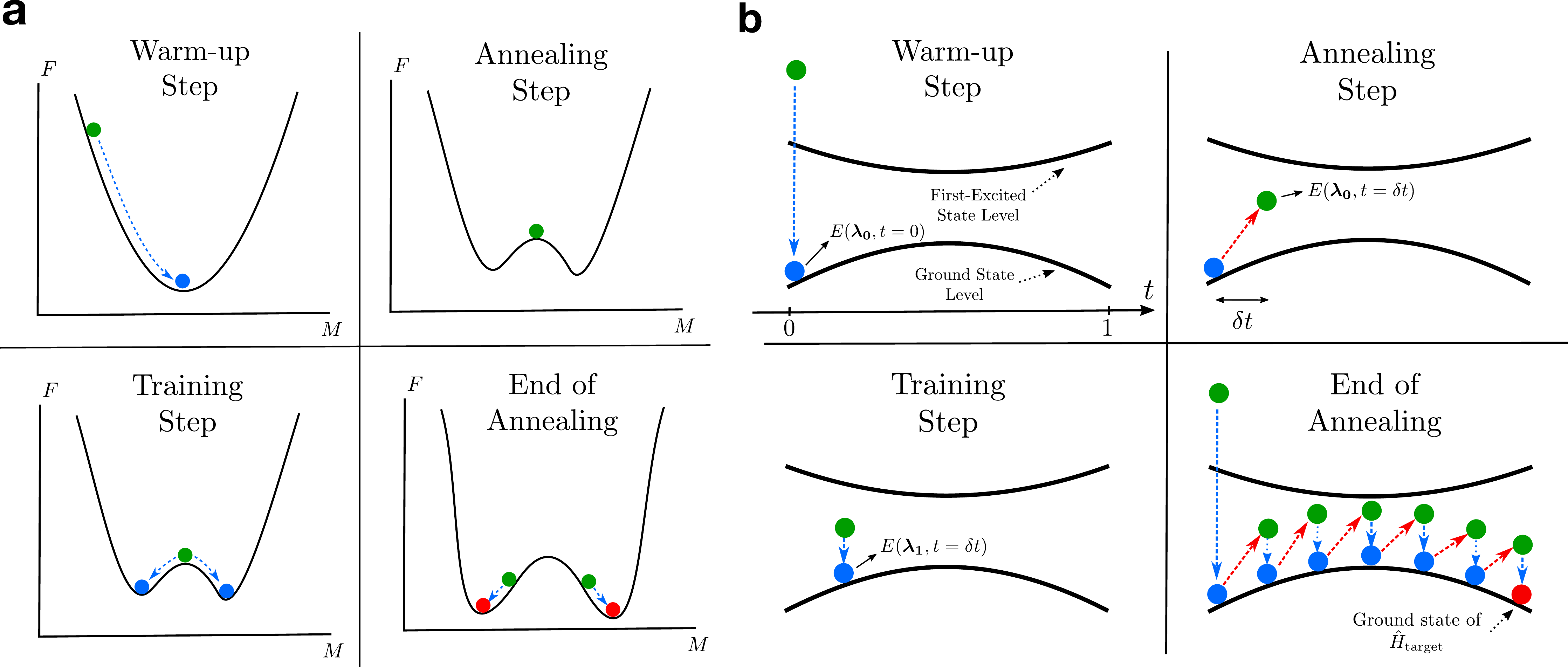}
    \caption{Variational neural annealing protocols. (a) The variational classical annealing (VCA) algorithm steps. A warm-up step brings the initialized variational state (green dot) close to the minimum of the free energy (cyan dot) at a given value of the order parameter $M$. This step is followed by an annealing and a training step that brings the variational state back to the new free energy minimum. Repeating the last two steps until $T(t=1)=0$ (red dots) produces approximate solutions to $H_{\rm target}$  if the protocol is conducted slowly enough. This schematic illustration corresponds to annealing through a continuous phase transition with an order parameter $M$.  (b) Variational quantum annealing (VQA). VQA includes a warm-up step, followed by an annealing and a training step, which brings the variational energy (green dot) closer to the new a ground state energy (cyan dot). We loop over the previous two steps until reaching the target ground state of $\hat{H}_{\rm target}$ (red dot) if annealing is performed slowly enough. }
    \label{fig:variationalclassicalannealing}
\end{figure*}

We first consider the variational approach to statistical mechanics~\cite{feynman1998statistical,Wu_2019}, where a distribution $p_{\bm{\lambda}}(\bm{\sigma})$ defined by a set of variational parameters $\bm{\lambda}$ is optimized to closely reproduce the equilibrium properties of a system at temperature $T$. Following the spirit of SA, 
we dub our first variational neural annealing algorithm {\it variational classical annealing} (VCA).

The VCA algorithm searches for the ground state of an optimization problem, encoded in a target Hamiltonian $H_{\rm target}$, by slowly annealing the model's variational free energy   
\begin{equation}
    F_{\bm{\lambda}}(t) = \langle H_{\rm target} \rangle_{\bm{\lambda}} - T(t) S_{\rm classical} ( p_{\bm{\lambda}} ),
    \label{eq:FreeEnergy}
\end{equation}
from a high temperature to a low temperature. The quantity $F_{\bm{\lambda}}(t)$ provides an upper bound to the true instantaneous free energy and can be used at each annealing stage to update $\bm{\lambda}$ through gradient-descent techniques. The brackets $\langle ... \rangle_{\bm{\lambda}}$ denote ensemble averages taken over the probability $p_{\bm{\lambda}}(\bm{\sigma})$.  The von Neumann entropy is given by
\begin{equation}
    S_{\rm classical} (p_{\bm{\lambda}}) = - \sum_{\bm{\sigma}} p_{\bm{\lambda}}(\bm{\sigma}) \log\left(p_{\bm{\lambda}}(\bm{\sigma})\right),
    \label{eq:vnEntropy}
\end{equation}
where the sum runs over all the elements of the state space $\{\bm{\sigma}\}$. In our setting, the temperature is decreased from an initial value $T_0$ to $0$ using a linear schedule function $T(t)=T_0(1-t)$, where $t\in [0,1]$, which follows closely the traditional implementation of SA.

In order for VCA to succeed, we require parameterized models that enable the estimation of entropy, Eq.~\eqref{eq:vnEntropy}, without incurring expensive calculations of the partition function. In addition, we anticipate that hard optimization problems will induce a complex energy landscape into the parameterized models and an ensuing slowdown of their sampling via Markov chain Monte Carlo. These issues preclude un-normalized models such as restricted Boltzmann machines, where sampling relies on Markov chains and whose partition function is intractable to evaluate~\cite{10.5555/3104322.3104412}. Instead, we implement VCA using recurrent neural networks (RNNs)~\cite{RNNWF_2020, roth2020iterative}, whose autoregressive nature enables statistical averages over exact samples $\bm{\sigma}$ drawn from $p_{\bm{\lambda}}(\bm{\sigma})$. Since RNNs are normalized by construction, these samples naturally allow the estimation of the entropy in Eq.~\eqref{eq:vnEntropy}. We provide a detailed description of the RNN in Methods Sec.~\ref{sec:RNNWF}.

The VCA algorithm, summarized in Fig.~\ref{fig:variationalclassicalannealing}(a), performs a warm-up step which brings a randomly initialized distribution $p_{\bm{\lambda}}(\bm{\sigma})$ to an approximate equilibrium state with free energy $F_{\bm{\lambda}}(t=0)$ via $N_{\rm warmup}$ gradient descent steps. At each step $t$, we reduce the temperature of the system from $T(t)$ to $T(t+\delta t)$ and apply $N_{\rm train}$ gradient descent steps to re-equilibrate the model. A critical ingredient to the success of VCA is that the variational parameters optimized at temperature $T(t)$ are reused at temperature $T(t +\delta t)$ to ensure that the model's distribution is always near its instantaneous equilibrium state. Repeating the last two steps $N_{\rm annealing}$ times, we reach temperature $T(1)=0$, which is the end of the annealing protocol. Here the distribution $p_{\bm{\lambda}}(\bm{\sigma})$ is expected to assign high probability to configurations $\bm{\sigma}$ that solve the optimization problem. Likewise, the residual entropy Eq.~\eqref{eq:vnEntropy} at $T(1)=0$ provides a heuristic approach to count the number of solutions to the problem Hamiltonian~\cite{Wu_2019}. Further algorithmic details are provided in Methods Sec.~\ref{app:VCA}. 



Simulated annealing provides a powerful heuristic for the solution of hard optimization problems by harnessing thermal fluctuations.  
Inspired by the latter, the advent of commercially available quantum devices~\cite{boixo2014evidence} has enabled the analogous
concept of quantum annealing \cite{Kadowaki_1998},
where the solution to an optimization problem is performed by harnessing quantum fluctuations.
In quantum annealing, the search for the ground state of Eq.~\eqref{eq:IsingGlassHamiltonian} is performed at $T=0$, 
by supplementing the target Hamiltonian with a quantum mechanical kinetic (or ``driving'') term,
\begin{equation}
    \hat{H}(t) = \hat{H}_{\text{target}}  + f(t)\hat{H}_D,
    \label{eq:qaHamiltonian}
\end{equation}
where $H_{\text{target}}$ in Eq.~\eqref{eq:IsingGlassHamiltonian} is promoted to a quantum mechanical Hamiltonian $\hat{H}_{\text{target}}$. 

Quantum annealing algorithms typically start with a dominant driving term $\hat{H}_D \gg \hat{H}_{\text{target}}$ chosen so that the ground state of $\hat{H}(0)$ is easy to prepare. When the strength of the driving term is subsequently reduced (typically adiabatically) using a schedule function $f(t)$, the system is annealed to the ground state of $\hat{H}_{\text{target}}$.
In analogy to its thermal counterpart, SQA emulates this process on classical computers using quantum Monte Carlo methods~\cite{Santoro_2002}.

Here, we leverage the variational principle of quantum mechanics and devise a strategy that emulates quantum annealing variationally.
We dub our second variational neural annealing algorithm {\it variational quantum annealing} (VQA). The latter is based on the variational Monte Carlo (VMC) algorithm, whose goal is to simulate the equilibrium properties of quantum systems at zero temperature (see Methods Sec.~\ref{app:VMC}). In VMC, the ground state of a Hamiltonian $\hat{H}$ is modeled through an ansatz $\ket{\Psi_{\bm{\lambda}}}$ endowed with parameters $\bm{\lambda}$. The variational principle guarantees that the energy $\langle \Psi_{\bm{\lambda}} |\hat{H} |\Psi_{\bm{\lambda}} \rangle$ is an upper bound to the ground state energy of $\hat{H}$, which we use to define a time-dependent objective function $ E(\bm{\lambda}, t) \equiv \langle  \hat{H}(t) \rangle_{\bm{\lambda}}=\langle \Psi_{\bm{\lambda}} |\hat{H}(t) |\Psi_{\bm{\lambda}} \rangle$ to optimize the parameters $\bm{\lambda}$.

The VQA setup, graphically summarized in Fig.~\ref{fig:variationalclassicalannealing}(b), applies $N_{\rm warmup}$ gradient descent steps to minimize $ E(\bm{\lambda}, t=0)$, which brings $\ket{\Psi_{\bm{\lambda}}}$ close to the ground state of $\hat{H}(0)$. Setting $t = \delta t$  while keeping the parameters $\bm{\lambda}_0$ fixed results in a variational energy $E(\bm{\lambda}_0, t=\delta t)$. A set of $N_{\rm train}$ gradient descent steps bring the ansatz closer to the new instantaneous ground state, which results in a variational energy $E(\bm{\lambda}_1, t=\delta t)$. The variational parameters optimized at time step $t$ are reused at time $t +\delta t$, which promotes the computational adiabaticity of the protocol (see Appendix.~\ref{app:adiabaticity_proof}). 
We repeat the annealing and training steps $N_{\rm annealing}$ times on a linear schedule ($f(t) = 1-t$ with $t \in [0,1]$) until $t = 1$, at which point the system should solve the optimization problem (red dot in Fig.~\ref{fig:variationalclassicalannealing}(b)). We note that in our simulations, no training steps are taken at $t=1$. 
Finally, similarly to VCA, we choose normalized RNN wave functions~\cite{RNNWF_2020, roth2020iterative} as ans\"atze, giving the VQA algorithm access to exact Monte Carlo samples. 

To gain theoretical insight on the principles behind a successful VQA simulation, we derive a variational version of the adiabatic theorem~\cite{Born1928}. Starting from a set of assumptions, such as the convexity of the energy landscape in the warm-up phase and close to convergence during annealing, as well as the absence of noise in the energy gradients, we provide a bound on the total number of gradient descent steps $N_{\text{steps}}$ that guarantees the adiabaticity of the VQA algorithm as well as a success probability of solving the optimization problem $P_{\rm success} > 1-\epsilon$. Here, $\epsilon$ is an upper bound on the overlap between the variational wave function and the excited states of the Hamiltonian $\hat{H}(t)$, i.e., $|\langle \Psi_{\perp}(t) | \Psi_{\bm{\lambda}} \rangle|^2 < \epsilon$. We show  that $N_{\text{steps}}$ can be bounded as (see Appendix.~\ref{app:derivation_annealingtime}):
\begin{equation}
     \mathcal{O} \left(\frac{\text{poly}(N)}{\epsilon \displaystyle\min_{\{t_n\}}(g(t_n))} \right) \leq N_{\text{steps}} \leq \mathcal{O} \left(\frac{\text{poly}(N)}{\epsilon^2 \displaystyle\min_{\{t_n\}}(g(t_n))^2} \right).
    \label{eq:num_steps}
\end{equation}
The function $g(t)$ is the energy gap between the first excited state and the ground state of the instantaneous Hamiltonian $\hat{H}(t)$, $N$ is the system size, and the set of times $\{t_n\}$ is defined in Appendix.~\ref{app:derivation_annealingtime}. As expected for hard optimization problems, the minimum gap typically decreases exponentially with system size $N$, which dominates the computational complexity of a VQA simulation, but in cases where the minimum gap scales as the inverse of a polynomial in $N$, then the number of steps $N_{\rm steps}$ is also polynomial in $N$. 

\section{Results}
\label{sec:results}
\subsection{Annealing on random Ising chains}
\label{sec:random_ising_chains}
We now proceed to evaluate the power of VCA and VQA. As a first benchmark, we consider the task of solving for the ground state the one-dimensional (1D) Ising Hamiltonian with random couplings $J_{i,i+1}$, 
 \begin{equation}
     H_{\text{target}} = -\sum_{i=1}^{N-1} J_{i,i+1} \sigma_i \sigma_{i+1}.
 \end{equation}

First, we examine $J_{i,i+1}$ sampled from a uniform distribution in the interval $[0,1)$. Here, the ground state configuration is given either by all spins up or down, and the ground state energy is known exactly, i.e., $E_G = -\sum_{i=1}^{N-1} J_{i,i+1}$~\cite{PhysRevB.99.064201}. 

We use a tensorized RNN ansatz without weight sharing for both VCA and VQA (see Methods Sec.~\ref{sec:RNNWF}). We consider system sizes $N=32,64,128$ and $N_{\rm train} = 5$, which suffices to achieve accurate solutions. For VQA, we use a one-body driving term $\hat{H}_D= -\Gamma_0  \sum_{i = 1}^{N} \hat{\sigma}_i^{x}$,  where $ \hat{\sigma}_i^{x,y,z}$ are Pauli matrices acting on site $i$. To quantify the performance of the algorithms, we use the residual energy~\cite{Santoro_2002}, 
\begin{equation}\label{eq:resen}
    \epsilon_{\text{res}} = {\big[ \langle H_{\rm target} \rangle_{\text{av}} - E_{\text{G}} \big]}_{\text{dis}},
\end{equation}
where $E_{\text{G}}$ is the exact ground state energy of $H_{\text{target}}$. We use the arithmetic mean for statistical averages $ \langle \ldots \rangle_{\text{av}}$ over samples from the models. For VCA it means that $\langle H_{\rm target} \rangle_{\text{av}} \approx \langle H_{\rm target}  \rangle_{\bm{\lambda}}$, while for VQA the target Hamiltonian is promoted to $ \hat{H}_{\text{target}} = -\sum_{i=1}^{N-1} J_{i,i+1} \hat{\sigma}_i^{z} \hat{\sigma}_{i+1}^{z}$ and $\langle H_{\rm target} \rangle_{\text{av}} \approx \langle \hat{H}_{\rm target} \rangle_{\bm{\lambda}}$. We consider the typical (geometric) mean for averaging over instances of the target Hamiltonian, i.e., $ {\big[  ... \big]}_{\text{dis}} = \exp (\langle \ln( ... )\rangle_{\text{av}})$. The average in the argument of the exponential stands for arithmetic mean over different realizations of the couplings. We take advantage of the autoregressive nature of the RNN and sample $10^6$ configurations at the end of the annealing, which allows us to accurately estimate the model's arithmetic mean. The typical mean is taken over 25 instances of $ H_{\rm target}$. 

\begin{figure}
    \centering
    \includegraphics[width =\linewidth]{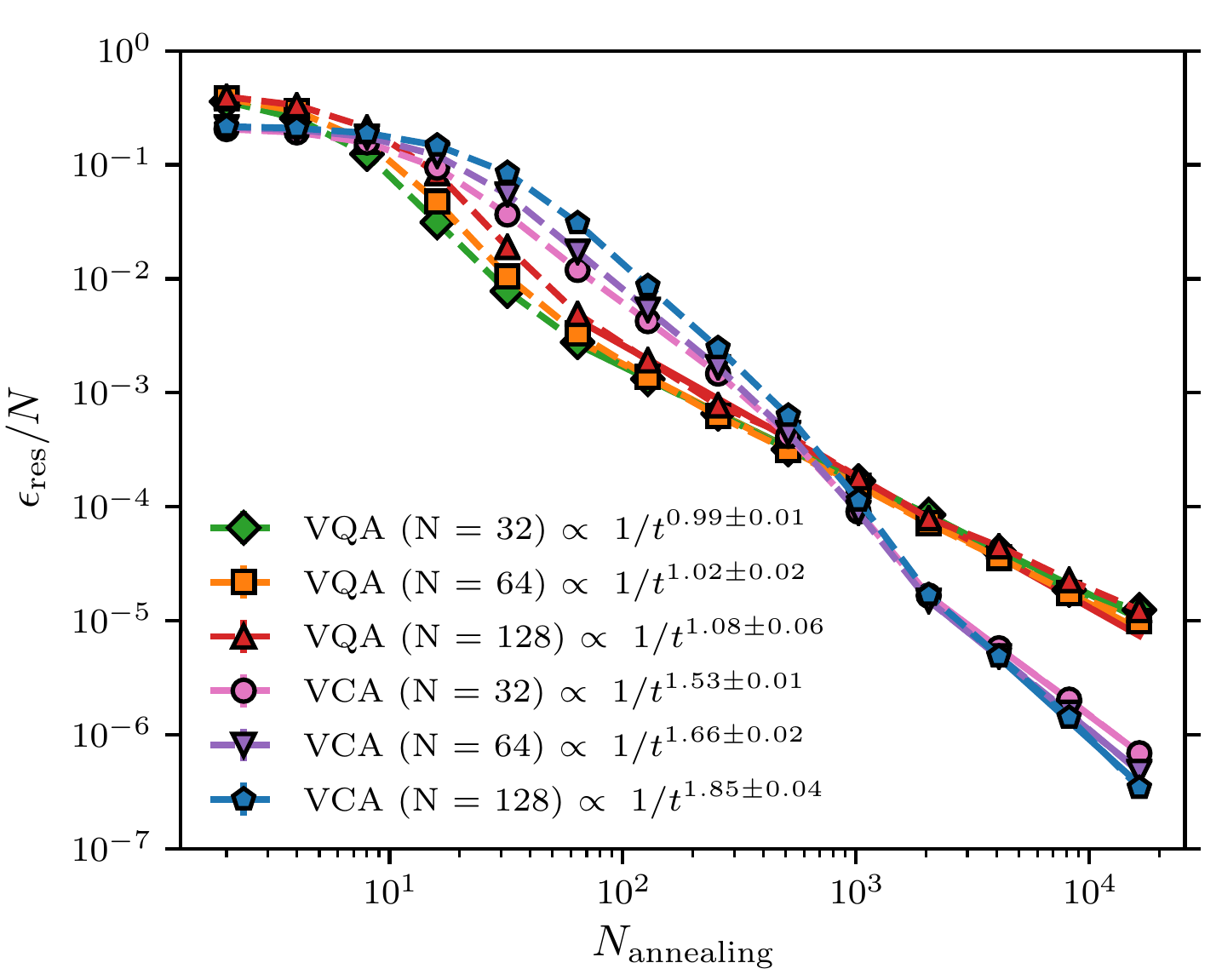}
    \caption{Variational neural annealing on a random Ising chain. Here we represent the residual energy per site $\epsilon_{\rm res}/N$ vs the number of annealing steps $N_{\rm annealing}$ for both VQA and VCA. The system sizes are $N = 32,64,128$. We use random positive couplings $J_{i,i+1} \in [0,1)$ (see text for more details). The error bars represent the one s.d. statistical uncertainty calculated over different disorder realizations~\cite{Norris1940}.
    } 
    \label{fig:1Dscaling}
\end{figure}

In Fig.~\ref{fig:1Dscaling} we report the residual energies per site against the number of annealing steps $N_{\rm annealing}$. As expected, the residual energy is a decreasing function of $N_{\rm annealing}$, which underlines the importance of adiabaticity and annealing in our setting. 
In our examples, we observe that the decrease of the residual energy of VCA and VQA is consistent with a power-law decay for a large number of annealing steps. Whereas VCA's decay exponent is in the interval $1.5-1.9$,  the VQA exponent is about $0.9-1.1$. These exponents suggest
an asymptotic speed-up compared to SA and coherent quantum annealing, where the residual energies follow a logarithmic law~\cite{PhysRevB.93.224431}. Contrary to the observations in Ref.~\cite{PhysRevB.93.224431} where quantum annealing was found superior to SA, VCA finds an average residual energy an order of magnitude more accurate than VQA for a large number of annealing steps.

Finally, we note that the exponents provided above are not expected to be universal and are a priori sensitive to the hyperparameters of the algorithms, e.g., learning rate, model choice, number of training steps, optimizer, etc. Appendix.~\ref{app:hyperparameters} provides a summary of the hyperparameters used in our work. Additional illustrations of the adiabaticity of VCA and VQA, as well as of the annealing results for a chain with $J_{i,i+1}$ uniformly sampled from the discrete set $\{-1,+1\}$, are provided in Appendix.~\ref{app:adiabaticity_proof}.

\subsection{Edwards-Anderson model}
\label{sec:EA}

We now consider the two-dimensional (2D) Edwards-Anderson (EA) model, which is a prototypical spin glass arranged on a square lattice with nearest neighbor random interactions. The problem of finding ground states of the model has been studied experimentally~\cite{Brooke1999} and numerically~\cite{Santoro_2002} from the annealing perspective, as well as theoretically~\cite{Barahona_1982} from the computational complexity perspective. The EA model with open boundary conditions is given by 
\begin{equation}
     H_{\rm target} = -\sum_{\langle i, j \rangle} J_{ij} \sigma_i\sigma_{j},
\end{equation}
where $\langle i, j \rangle$ denote nearest neighbors. The couplings $J_{ij}$ are drawn from a uniform distribution in the interval $\left[-1,1\right)$. In the absence of a longitudinal field, for which solving the EA model is NP-hard, the ground state can be found in polynomial time~\cite{Barahona_1982}. To find the exact ground state of each random realization, we use the spin-glass server~\cite{spinglass_server}. 
\begin{figure}
    \centering
    \includegraphics[width =\linewidth]{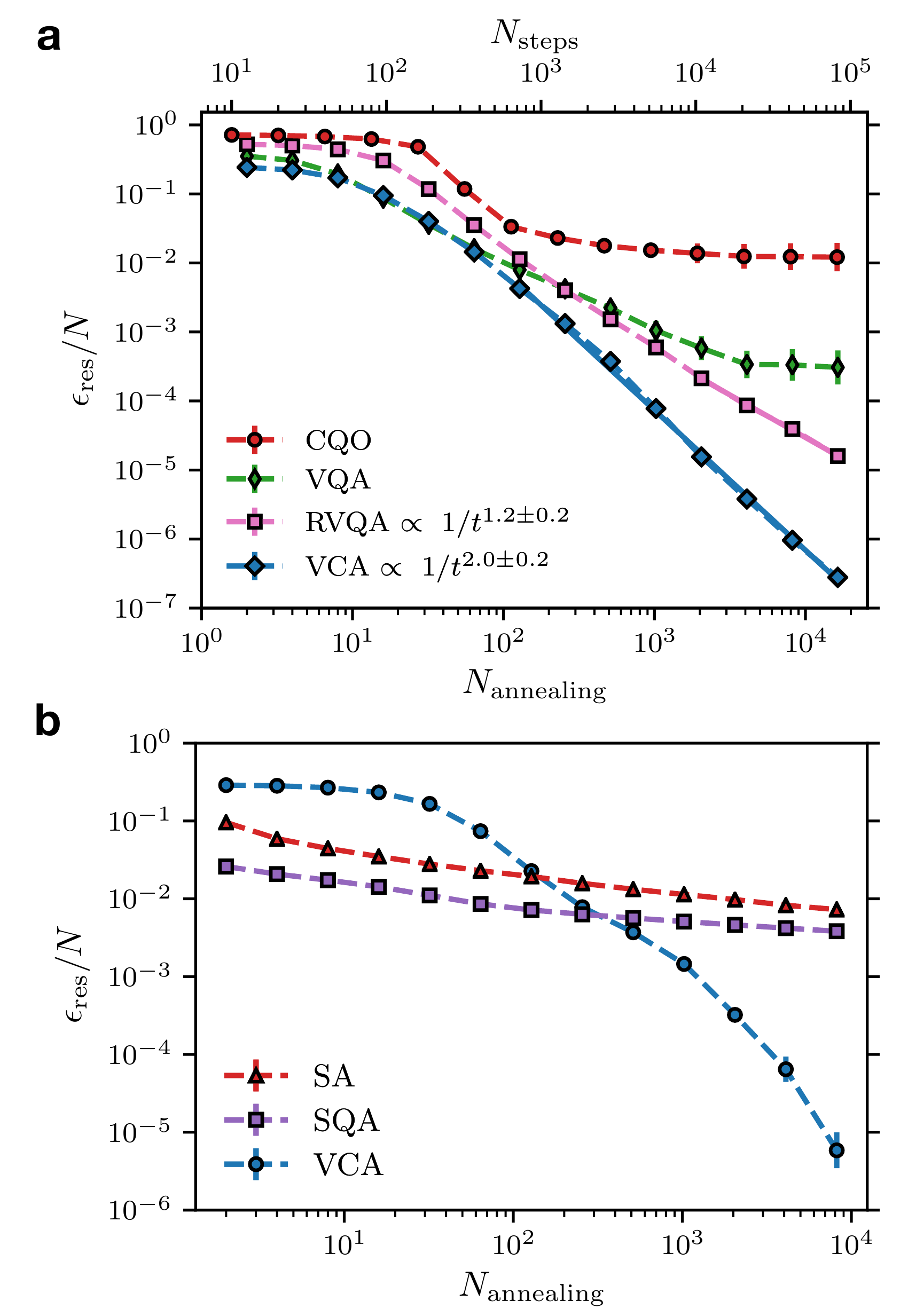} 
    \caption{Benchmarking the two-dimensional Edwards-Anderson spin glass. (a) A comparison between VCA, VQA, RVQA, and CQO  on a $10 \times 10$ lattice by plotting the residual energy per site vs $N_{\text{annealing}}$. For CQO, we report the residual energy per site vs the number of optimization steps $N_{\rm steps}$. (b) Comparison between SA, SQA with $P=20$ trotter slices, and VCA using a 2D tensorized RNN ansatz on a $40 \times 40$ lattice. The annealing speed is the same for SA, SQA and VCA. } 
    \label{fig:EA_comaprison} 
\end{figure}

We use a 2D tensorized RNN ansatz without weight sharing for the variational protocols (see Methods Sec.~\ref{sec:RNNWF}). For VQA, we use a one-body driving term $\hat{H}_D= -\Gamma_0  \sum_{i = 1}^{N} \hat{\sigma}_i^{x}$. Fig.~\ref{fig:EA_comaprison}(a) shows the annealing results obtained on a system size $N=10\times10$ spins. VCA outperforms VQA and in the adiabatic, long-time annealing regime, it produces solutions three orders of magnitude more accurate on average than VQA. In addition, we investigate the performance of VQA supplemented with a fictitious Shannon information entropy~\cite{roth2020iterative} term that mimics thermal relaxation effects observed in quantum annealing hardware~\cite{dickson2013thermally}. 
This form of regularized VQA, here labelled (RVQA), is described by a pseudo free energy cost function $\tilde{F}_{\bm{\lambda}}(t)=\langle \hat{H}(t) \rangle_{\bm{\lambda}}- T(t) S_{\rm classical} (|\Psi_{\bm{\lambda}}|^2)$. As in VCA, the pseudo entropy term $S_{\rm classical} (|\Psi_{\bm{\lambda}}|^2)$ at $f(1)=0$ provides a heuristic approach to count the number of solutions to $H_{\rm target}$ for VQA and RVQA. The results in Fig.~\ref{fig:EA_comaprison}(a) do show an amelioration of the VQA performance, including changing a saturating dynamics at large $N_{\rm annealing}$ to a power-law like behavior. However, it appears to be insufficient to compete with the VCA scaling (see exponents in Fig.~\ref{fig:EA_comaprison}(a)). This observation suggests the superiority of a thermally driven variational emulation of annealing over a purely quantum one for this example. 

To further scrutinize the relevance of the annealing effects in VCA, we also consider VCA with zero thermal fluctuations, i.e., setting $T_0 = 0$. Because of its intimate relation to the classical-quantum optimization (CQO) methods of Refs.~\cite{gomes2019classical,sinchenko2019deep, zhao2020natural}, we refer to this setting as CQO. Fig.~\ref{fig:EA_comaprison}(a) shows that CQO takes about $10^3$ training steps to reach accuracies nearing $1\%$. The accuracy does not further improve upon additional training up to $10^5$ gradient steps, which indicates that CQO is prone to getting stuck in local minima. In comparison, VCA and VQA offer solutions orders of magnitude more accurate on average for a large number of annealing steps, highlighting the importance of annealing in tackling optimization problems.

Since VCA displays the best performance in the previous benchmarks, we use it to demonstrate its capabilities on a $40 \times 40$ spin system. For comparison, we use SA as well as SQA. The SQA simulation uses the
path-integral Monte Carlo method~\cite{Santoro_2002} with $P=20$ trotter slices, and we report averages over energies across all trotter slices, for each realization of randomness (see Methods Sec.~\ref{app:PIQMC}). In addition, we average the energy obtained after $25$ annealing runs on every instance of randomness for SA and SQA. To average over Hamiltonian instances, we use the typical mean over $25$ different realizations for the three annealing methods. The results are shown in Fig.~\ref{fig:EA_comaprison}(b), where we present the residual energies per site against the number of annealing steps $N_{\rm annealing}$, which is set so that the speed of annealing is the same for SA, SQA and VCA. We first note that our results confirm the qualitative behavior of SA and SQA in Refs.~\cite{Santoro_2002, Santoro2002_2}. While SA and SQA produce lower residual energy solutions than VCA for small $N_{\rm annealing}$, we observe that VCA achieves residual energies about three orders of magnitude smaller than SQA and SA for a large number of annealing steps. Notably, the rate at which the residual energy improves with increasing $N_{\rm annealing}$ is significantly higher for VCA compared to SQA and SA even at relatively small number of annealing steps. 

\subsection{Fully-connected spin glasses}
\label{sec:fullyconnected_spin_glass}

\begin{figure*}
    \centering
    \includegraphics[width =\linewidth]{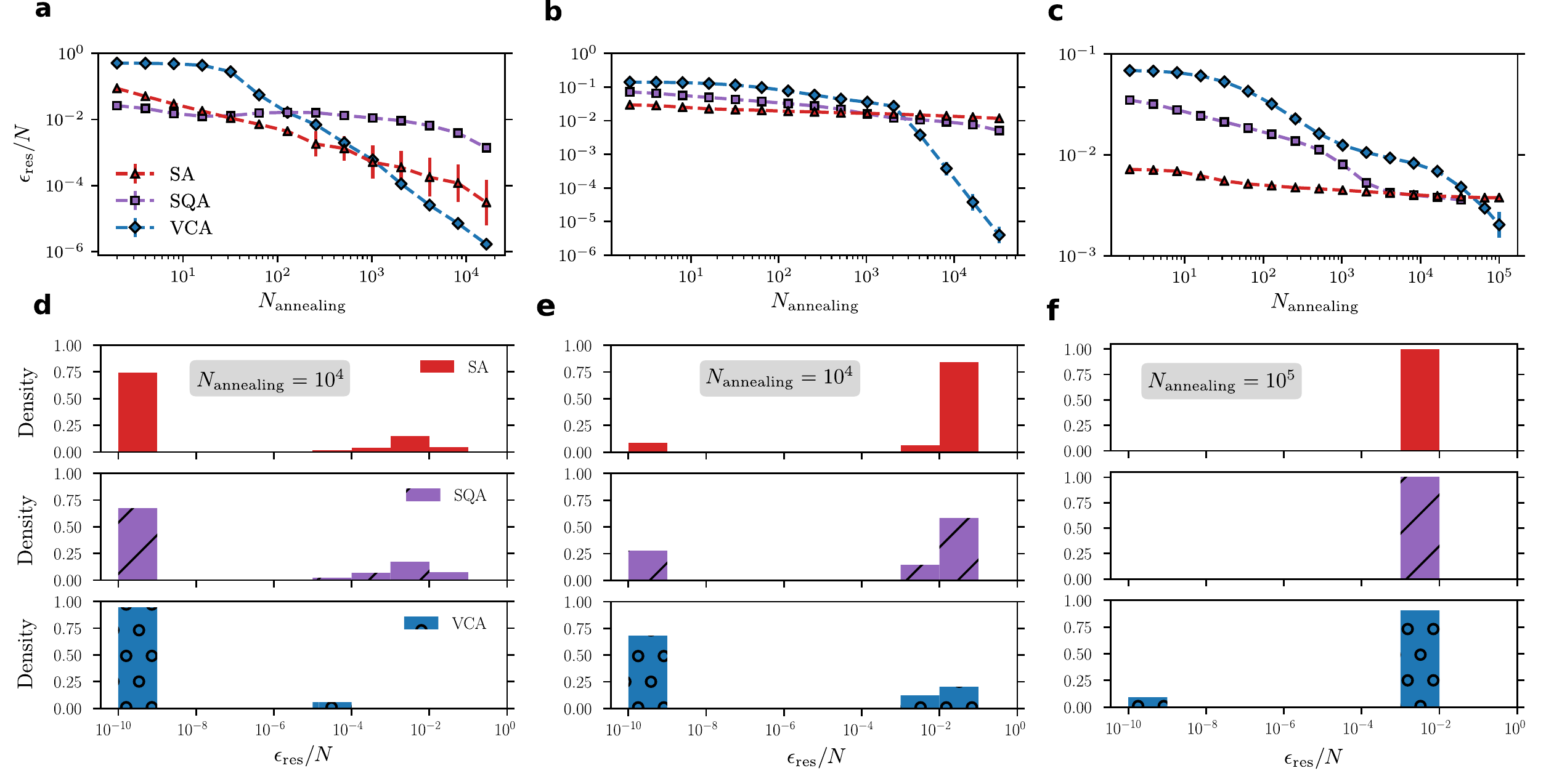}
  \caption{
Benchmarking SA, SQA ($P=100$ trotter slices) and VCA on the Sherrington-Kirkpatrick (SK) model and the Wishart planted ensemble (WPE). Panels (a),(b), and (c) display the residual energy per site as a function of $N_{\rm annealing}$.  (a) The SK model with $N = 100$ spins. (b) WPE with $N=32$ spins and $\alpha = 0.5$. (c) WPE with $N=32$ spins and $\alpha = 0.25$.  Panels (d), (e) and (f) display the residual energy histogram for each of the different techniques and models in panels (a),(b), and (c), respectively. The histograms use  $25000$ data points for each method. Note that we choose a minimum threshold of $10^{-10}$ for $\epsilon_{\rm res}/N$, which is within our numerical accuracy.}
    \label{fig:SAvsPIQMCvsRNN}
\end{figure*}

We now focus our attention on fully-connected spin glasses~\cite{Barahona_1982, Mezard1986}. We first focus on the Sherrington-Kirkpatrick (SK) model~\cite{PhysRevLett.35.1792}, which provides a conceptual framework for the understanding of the role of disorder and frustration in widely diverse systems ranging from materials to combinatorial optimization and machine learning. The SK Hamiltonian is given by
\begin{equation}
    H_{\text{target}} = -\frac{1}{2} \sum_{i \neq j} \frac{J_{ij}}{\sqrt{N}} \sigma_i \sigma_j,
\end{equation}
where $\{J_{ij}\}$ is a symmetric matrix such that each matrix element $J_{ij}$ is sampled from a gaussian distribution with mean $0$ and variance $1$.

Since VCA performed best in our previous examples, we use it to find ground states of the SK model for $N = 100$ spins. Here, exact ground states energies of the SK model are calculated using the spin-glass server~\cite{spinglass_server} on a total of $25$ instances of disorder. To account for long-distance dependencies between spins in the SK model, we use a dilated RNN ansatz that has $\lceil \log_2(N) \rceil = 7$ layers (see Methods Sec.~\ref{sec:RNNWF}) and set the initial temperature $T_0 = 2$. We compare our results with SA and SQA. For SQA, we start with an initial magnetic field $\Gamma_0 = 2$, while for SA we use $T_0 = 2$.  

For an effective comparison, we first plot the residual energy per site as a function of $N_{\rm annealing}$ for VCA, SA and SQA (with $P = 100$ trotter slices). Here, the SA and SQA residual energies are obtained by averaging the outcome of $50$ independent annealing runs, while for VCA we average the outcome of $10^6$ exact samples from the annealed RNN. For all methods, we take the typical average over 25 disorder instances. The results are shown in Fig.~\ref{fig:SAvsPIQMCvsRNN}(a). As observed in the EA model, we note that SA and SQA produce lower residual energy solutions than VCA for small $N_{\rm annealing}$, but we emphasize that VCA delivers a lower residual energy compared to SQA and SA as the total number of annealing steps increases past $N_{\rm annealing}\sim 10^{3}$. Likewise, we observe that the rate at which the residual energy improves with increasing $N_{\rm annealing}$ is significantly higher for VCA in comparison to SQA and SA.  

A more detailed look at the statistical behaviour of the methods at large $N_{\rm annealing}$ can be obtained from the residual energy histograms separately produced by each method, as shown in Fig.~\ref{fig:SAvsPIQMCvsRNN}(d). The histograms contain $1000$ residual energies for each of the same $25$ disorder realizations. For each instance, we plot results for $1000$ SA runs, $1000$ samples obtained from the RNN at the end of annealing for VCA, and $10$ SQA runs including contribution from each of the $P=100$ Trotter slices. 
We observe that VCA is superior to SA and SQA, as it produces a higher density of low energy configurations. This indicates that, even though VCA typically takes more annealing steps, it ultimately results in a higher chance of getting more accurate solutions to optimization problems than SA and SQA. Note that for the SK model, the SQA histogram remain quantitatively the same for 200 runs, and we report data of 10 runs only for fairness purposes compared to both SA and VCA.

We now focus on the Wishart planted ensemble (WPE), which is a class of zero-field Ising models with a first-order phase transition and tunable algorithmic hardness~\cite{Hamze_2020}. These problems belong to a special class of hard problem ensembles whose solutions are known a priori, which, together with the tunability of the hardness, makes the WPE model an ideal tool to benchmark heuristic algorithms for optimization problems. The Hamiltonian of the WPE model is defined as
\begin{equation}
    H_{\text{target}} = -\frac{1}{2} \sum_{i \neq j} J^{\alpha}_{ij} \sigma_i \sigma_j.
\end{equation}
Here $J^{\alpha}_{ij}$ is a symmetric matrix satisfying 
\begin{equation*}
    J^{\alpha} =  \tilde{J}^{\alpha} - \text{diag}(\tilde{J})
\end{equation*}
and 
\begin{equation*}
    \tilde{J}^{\alpha} = -\frac{1}{N} W_{\alpha} W_{\alpha}^{\text{T}}.
\end{equation*}
The term $W_{\alpha}$ is an $N \times \lfloor \alpha N \rfloor$ random matrix satisfying $W_{\alpha}t_{\text{ferro}} = 0$ where $t_{\text{ferro}} = (+1,+1, ..., +1)$ is the ferromagnetic state (see Ref.~\cite{Hamze_2020} for details about the generation of $W_{\alpha}$).  The ground state of the WPE model is known (i.e., it is planted) and corresponds to the ferromagnetic states $\pm t_{\text{ferro}}$. Interestingly, $\alpha$ is a tunable parameter of hardness, where for $\alpha<1$ this model displays a first-order transition, such that near zero temperature the paramagnetic states are meta-stable solutions~\cite{Hamze_2020}. This feature makes this model hard to solve with any annealing method, as the paramagnetic states are numerous compared to the two ferromagnetic states and hence act as a trap for a typical annealing method. We benchmark the three methods (SA, SQA and VCA) for $N=32$ and $\alpha \in \{0.25, 0.5\}$. 

We consider $25$ instances of the couplings $\{J^{\alpha}_{ij}\}$ and attempt to solve the model with VCA implemented using a dilated RNN ansatz with $\lceil \log_2(N) \rceil = 5$ layers and an initial temperature $T_0 = 1$. For SQA ($P=100$ trotter slices), we use an initial magnetic field $\Gamma_0 = 1$, and for SA we start with $T_0 = 1$. 

We first plot the scaling of residual energies per site $\epsilon_{\rm res}/N$ as shown in Figs.~\ref{fig:SAvsPIQMCvsRNN}(b) and (c). Here we note that VCA is superior to SA and SQA for $\alpha = 0.5$ as demonstrated in Fig.~\ref{fig:SAvsPIQMCvsRNN}(b). More specifically, VCA is about three orders of magnitude more accurate than SQA and SA for a large number of annealing steps. In the case of $\alpha = 0.25$ in Fig.~\ref{fig:SAvsPIQMCvsRNN}(c), VCA is competitive where it achieves a similar performance compared to SA and SQA on average for a large number of annealing steps. We also represent the residual energies in a histogram form. We observe that for $\alpha = 0.5$ in Fig.~\ref{fig:SAvsPIQMCvsRNN}(e), VCA achieves a higher density toward low residual energies $ \epsilon_{\text{res}}/N \sim 10^{-9}$-$10^{-10}$ compared to SA and SQA. For $\alpha = 0.25$ in Fig.~\ref{fig:SAvsPIQMCvsRNN}(f), VCA leads to a non-negligible density at very low residual energies as opposed to SA and SQA, whose solutions display residual energies orders of magnitude higher. Finally, our WPE simulations support the observation that VCA tends to improve the quality of solutions faster than SQA and SA for a large number of annealing steps.  

\section{Conclusions and outlook}

In conclusion, we have introduced a strategy to combat the slow sampling dynamics encountered by simulated annealing when an optimization landscape is rough or glassy.  Based on annealing the variational parameters of a generalized target distribution, our scheme — which we dub {\it variational neural annealing} — takes advantage of the power of modern autoregressive models, which can be exactly sampled without slow dynamics even when a rough landscape is encountered. We implement variational neural annealing parameterized by a recurrent neural network, and compare its performance to conventional simulated annealing on prototypical spin glass Hamiltonians known to have landscapes of varying roughness. We find that variational neural annealing produces accurate solutions to all of the optimization problems considered, including spin glass Hamiltonians where our techniques typically reach solutions orders of magnitude more accurate on average than conventional simulated annealing in the limit of a large number of annealing steps.

We emphasize that several hyperparameters, model, hardware, and variational objective function choices can be explored and may improve our methodologies. We have utilized a simple annealing schedule in our protocols and highlight that reinforcement learning can be used to improve it~\cite{mills2020controlled}. A critical insight gleaned from our experiments is that certain neural network architectures were more efficient on specific Hamiltonians. Thus, a natural direction is to study the intimate relation between the model architecture and the problem Hamiltonian, where we envision that symmetries and domain knowledge would guide the design of models and algorithms.

As we witness the unfolding of a new age for optimization powered by deep learning~\cite{BENGIO2020}, we anticipate a rapid adoption of machine learning techniques in the space of combinatorial optimization, as well as anticipate domain-specific applications of our ideas in diverse technological and scientific areas related to physics, biology, health care, economy, transportation, manufacturing, supply chain, hardware design, computing and information technology, among others.

\section{Methods}

\subsection{Recurrent Neural Network Ans\"atze}
\label{sec:RNNWF}

\begin{figure}
    \centering
    \includegraphics[width=\linewidth]{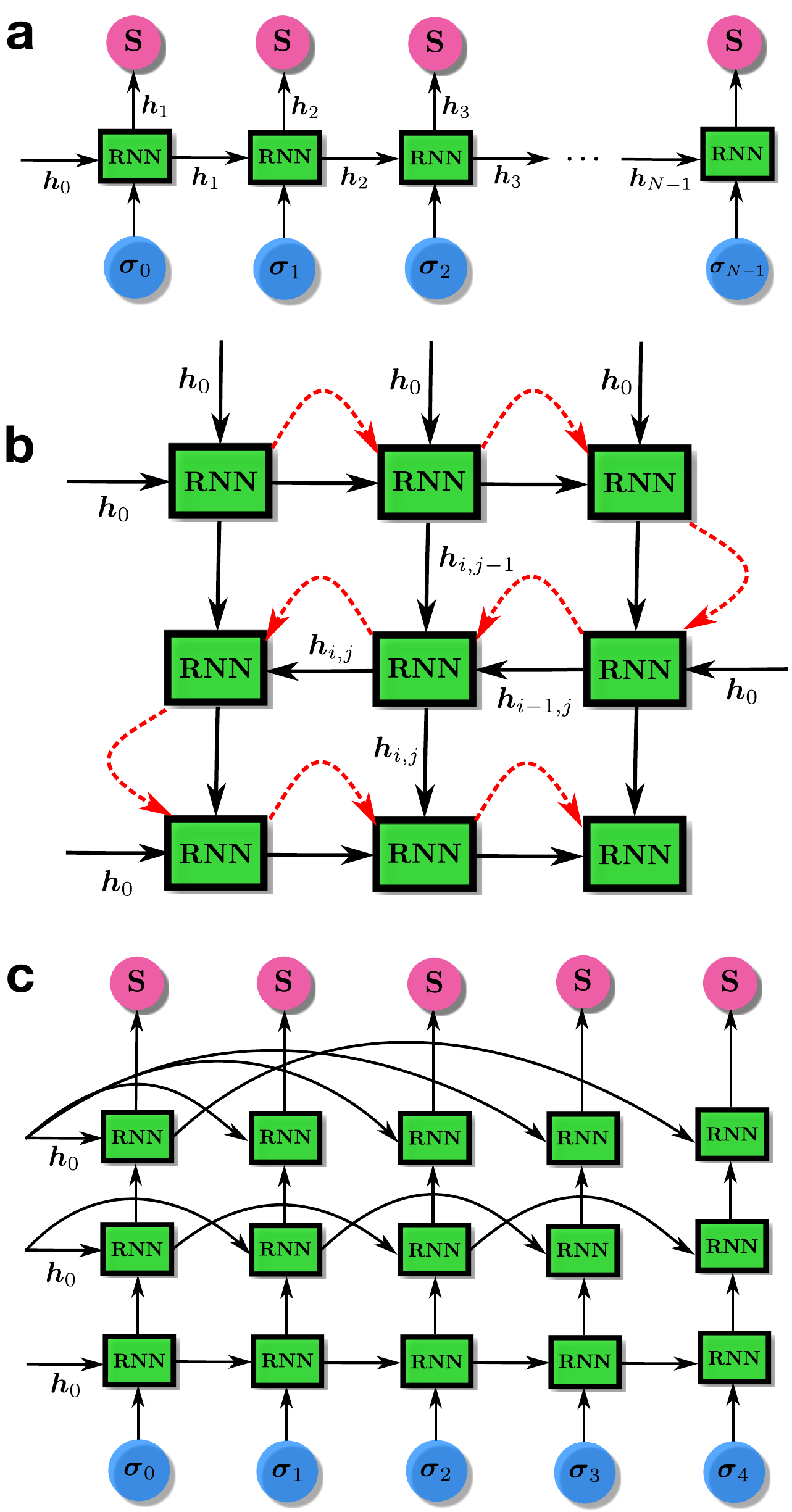}
    \caption{(a) An illustration of a 1D RNN: at each site $n$, the RNN cell denoted by the green box, receives a hidden state $\bm{h}_{n-1}$ and the one-hot spin vector $\bm{\sigma}_{n-1}$, to generate a new hidden state $\bm{h}_{n}$ that is fed into a Softmax layer (denoted by a magenta circle). (b) A graphical illustration of a 2D RNN. Each RNN cell receives two hidden states $\bm{h}_{i,j-1}$ and $\bm{h}_{i-1,j}$, as well as two input vectors $\bm{\sigma}_{i,j-1}$ and $\bm{\sigma}_{i-1,j}$ (not shown) as illustrated by the black arrows. The red arrows correspond to the zigzag path we use for 2D autoregressive sampling. The initial memory state $\bm{h}_0$ of the RNN and the initial inputs $\bm{\sigma}_0$ (not shown) are null vectors. (c) An illustration of a dilated RNN, where the distance between each two RNN cells grows exponentially with depth to account for long-term dependencies. We choose depth $L = \lceil{\log_2(N)}\rceil$ where $N$ is the number of spins.}
    \label{fig:RNN}
\end{figure}

Recurrent neural networks model complex probability distributions $p$ by taking advantage of the chain rule
\begin{align}
    p(\bm{ \sigma})= p(\sigma_1)p(\sigma_2|\sigma_1) \cdots p(\sigma_N|\sigma_{N-1}, \dots, \sigma_2, \sigma_1),
    \label{eq:prod_rule}
\end{align}
where specifying every conditional probability $p(\sigma_i| \sigma_{<i})$ provides a full characterization of the joint distribution $p(\bm{\sigma})$. Here, $\{\sigma_{n}\}$ are $N$ binary variables such that $\sigma_{n} = 0$ corresponds to a spin down while $\sigma_{n} = 1$ corresponds to a spin up. RNNs consist of elementary cells that parameterize the conditional probabilities. In their original form,  ``vanilla'' RNN cells~\cite{Goodfellow-et-al-2016} compute a new ``hidden state'' $\bm{h}_n$ with dimension $d_h$, for each site $n$, following the relation 
\begin{equation}
    \bm{h}_n = F(W [\bm{h}_{n-1}; \bm{\sigma}_{n-1}]+\bm{b}),
    \label{eq:vanillaRecursion}
\end{equation}
where $[\bm{h}_{n-1}; \bm{\sigma}_{n-1}]$ is vector concatenation of $\bm{h}_{n-1}$ and a one-hot encoding $\bm{\sigma}_{n-1}$ of the binary variable $\sigma_{n-1}$~\cite{RNNWF_2020}. The function $F$ is a non-linear activation function.  From this recursion relation, it is clear that the hidden state $\bm{h}_{n}$ encodes information about the previous spins $\bm{\sigma}_{n'<n}$. Hence, the hidden state $\bm{h}_n$ provides a simple strategy to model the conditional probability $p_{\bm{\lambda}}(\sigma_n| \sigma_{<n})$ as
\begin{equation}
    p_{\bm{\lambda}}(\sigma_n| \sigma_{<n}) = \text{Softmax}(U \bm{h}_n + \bm{c}) \cdot \bm{\sigma}_n,
    \label{eq:softmax_layer}
\end{equation}
where $\cdot$ denotes the dot product operation (see Fig.~\ref{fig:RNN}(a)). The set of all variational parameters of the model $\bm{\lambda}$ corresponds to $U, W,\bm{b},\bm{c}$, and 
\[
\text{Softmax}(\bm{v})_n = \frac{\exp(v_n)}{\sum_i \exp(v_i)}.
\]
The joint probability distribution $p_{\bm{\lambda}}(\bm{\sigma})$ is given by
\begin{equation}
    p_{\bm{\lambda}}(\bm{\sigma}) = p_{\bm{\lambda}}(\sigma_1)p_{\bm{\lambda}}(\sigma_2|\sigma_1) \cdots p_{\bm{\lambda}}(\sigma_N|\sigma_{<N}).
    \label{eq:RNN_prob}
\end{equation}
Since the outputs of the Softmax activation function sum to one, each conditional probability $p_{\bm{\lambda}}(\sigma_i|\sigma_{<i})$ is normalized, and hence $p_{\bm{\lambda}}(\bm{\sigma})$ is also normalized.

For disordered systems, it is natural to forgo the common practice of weight sharing~\cite{Goodfellow-et-al-2016} of $W,U, \bm{b}$ and $\bm{c}$ in Eqs.~\eqref{eq:vanillaRecursion}, \eqref{eq:softmax_layer} and use an extended set of site-dependent variational parameters $\bm{\lambda}$ comprised of $\{W_n\}_{n=1}^{N}$ and $\{U_n\}_{n=1}^{N}$ and biases $\{\bm{b}_n\}_{n=1}^{N}$, $\{\bm{c}_n\}_{n=1}^{N}$. The recursion relation and the Softmax layer are modified to 
\begin{equation}
    \bm{h}_n = F(W_n [\bm{h}_{n-1}; \bm{\sigma}_{n-1}]+\bm{b}_n),
    \label{eq:vanilla_NoWeightSharing}
\end{equation}
and 
\begin{equation}
    p_{\bm{\lambda}}(\sigma_n| \sigma_{<n}) = \text{Softmax}(U_n \bm{h}_n + \bm{c}_n) \cdot \bm{\sigma}_n,
    \label{eq:softmax_noweightsharing}
\end{equation}
respectively. Note that the advantage of not using weight sharing for disordered systems is further demonstrated in Appendix.~\ref{app:benchmarkingRNNcells}.

We also consider a tensorized version of vanilla RNNs which replaces the concatenation operation in Eq.~\eqref{eq:vanilla_NoWeightSharing} with the operation~\cite{kelley_2016}
\begin{equation}
    \bm{h}_{n} = F\left(\bm{\sigma}^{ \intercal }_{n-1} T_n 
    \bm{h}_{n-1} + \bm{b}_n \right),
    \label{eq:TRNN_recursion}
\end{equation}
where $\bm{\sigma}^{\intercal}$ is the transpose of $\bm{\sigma}$, and the variational parameters $\bm{\lambda}$ are $\{T_n\}_{n=1}^{N}$, $\{U_n\}_{n=1}^{N}$, $\{\bm{b}_n\}_{n=1}^{N}$ and $\{\bm{c}_n\}_{n=1}^{N}$. This form of tensorized RNN increases the expressiveness of our ansatz as illustrated in Appendix.~\ref{app:benchmarkingRNNcells}.

For two-dimensional systems, we make use of a 2D-dimensional extension of the recursion relation in vanilla RNNs~\cite{RNNWF_2020}
\begin{equation}
    \bm{h}_{i,j} = F\!\left(W_{i,j}^{(h)} 
    [\bm{h}_{i-1,j} ; \bm{\sigma}_{i-1,j}] + W_{i,j}^{(v)} 
    [\bm{h}_{i,j-1} ; \bm{\sigma}_{i,j-1}] + \bm{b}_{i,j}\right).
    \label{eq:2DVRNN}
\end{equation}
To enhance the expressive power of the model, we promote the recursion relation to a tensorized form  
\begin{equation}
    \bm{h}_{i,j} = F\!\left(
    [\bm{\sigma}_{i-1,j} ; \bm{\sigma}_{i,j-1}]T_{i,j} 
    [\bm{h}_{i-1,j} ; \bm{h}_{i,j-1}] + \bm{b}_{i,j}\right).
    \label{eq:2DTRNN}
\end{equation}
Here, $T_{i,j}$ are site-dependent weight tensors that have dimension $4 \times 2 d_h \times d_h$. We also note that the coordinates $(i-1, j)$ and $(i,j-1)$ are path-dependent, and are given by the zigzag path, illustrated by the black arrows in Fig.~\ref{fig:RNN}(b). Moreover, to sample configurations from the 2D tensorized RNNs, we use the same zigzag path as illustrated by the red dashed arrows in Fig.~\ref{fig:RNN}(b). 

For models such as the Sherrington-Kirkpatrick model and the Wishart planted ensemble, every spin interacts with each other. To account for the long-distance nature of the correlations induced by these interactions, we use dilated RNNs~\cite{chang2017dilated}, which are known to alleviate the vanishing gradient problem~\cite{Bengio1994}. Dilated RNNs are multi-layered RNNs that use dilated connections between spins to model long-term dependencies~\cite{NIPS1995_1102}, as illustrated in Fig.~\ref{fig:RNN}(c). At each layer $1\leq l\leq L$, the hidden state is computed as
\begin{equation*}
    \bm{h}^{(l)}_n = F(W^{(l)}_n [\bm{h}^{(l)}_{\text{max}(0, n-2^{l-1})}; \bm{h}^{(l-1)}_n] + \bm{b}^{(l)}_n).
\end{equation*}
Here $\bm{h}^{(0)}_n = \bm{\sigma}_{n-1}$ and the conditional probability is given by
\begin{equation*}
   p_{\bm{\lambda}}(\sigma_n|\sigma_{<n}) = \text{Softmax}(U_n \bm{h}^{(L)}_n + \bm{c}_n) \cdot \bm{\sigma}_n.
\end{equation*}
In our work, we choose the size of the hidden states $\bm{h}^{(l)}_n$, where $l>0$, as constant and equal to $d_h$. We also use a number of layers $L = \lceil{\log_2(N)}\rceil$, where $N$ is the number of spins and $\lceil{\ldots}\rceil$ is the ceiling function. This means that two spins are connected with a path whose length is bounded by $\mathcal{O}(\log_2(N))$, which follows the spirit of the multi-scale renormalization ansatz~\cite{Vidal_2008}. For more details on the advantage of dilated RNNs over tensorized RNNs see Appendix.~\ref{app:benchmarkingRNNcells}.

We finally note that for all the RNN architectures in our work, we found accurate results using the exponential linear unit (ELU) activation function, defined as:
 \begin{equation*}
     \text{ELU}(x) = \begin{cases}
 		  x, & \text{if } x \geq 0\,,  \\
 		\exp(x)-1, & \text{if } x < 0 \,.
 	\end{cases}
 \end{equation*}


\subsection{Minimizing the variational free energy}
\label{app:VCA}
To implement the variational classical annealing algorithm, we use the variational free energy 
\begin{equation}
    F_{\bm{\lambda}}(T) = \langle H_{\rm target} \rangle_{\bm{\lambda}} - T S_{\rm classical} ( p_{\bm{\lambda}} ),
    \label{eq:variational_freenergy}
\end{equation}
where the target Hamiltonian $H_{\rm target}$ encodes the optimization problem and $T$ is the temperature. Moreover, $S_{\text{classical}}$ is the entropy of the distribution $p_{\bm{\lambda}}$. To estimate $F_{\bm{\lambda}}(T)$ we take $N_s$ exact samples $\bm{\sigma}^{(i)} \sim p_{\bm{\lambda}}$  ($i=1,\ldots,N_s$) drawn from the RNN and evaluate
\begin{equation*}
     F_{\bm{\lambda}}(T) \approx \frac{1}{N_s}\sum_{i=1}^{N_s} F_{\rm loc}(\bm{\sigma}^{(i)}),
\end{equation*}
where the local free energy is $F_{\rm loc}(\bm{\sigma}) = H_{\rm target}(\bm{\sigma}) + T \log\left(p_{\bm{\lambda}}(\bm{\sigma})\right)$~\cite{Wu_2019}. Similarly, the gradients are given by 
\begin{align*}
    \partial_{\bm{\lambda}} F_{\bm{\lambda}}(T) \approx \frac{1}{N_s}\sum_{ i=1}^{N_s} & \partial_{\bm{\lambda}} \log \left(  p_{\bm{\lambda}}\left(\bm{\sigma}^{(i)}\right) \right)\\ 
    & \times \left(F_{\rm loc}(\bm{\sigma}^{(i)})  - F_{\bm{\lambda}}(T) \right),
\end{align*}
where we subtract $F_{\bm{\lambda}}(T)$ in order to reduce noise in the gradients~\cite{Wu_2019, RNNWF_2020}. We note that this variational scheme exhibits a zero-variance principle, namely that the local free energy variance per spin
\begin{equation}
    \sigma_F^2 \equiv \frac{\text{var}(\{F_{\rm loc}(\bm{\sigma})\})}{N},
    \label{eq:freeenergyvariance}
\end{equation}
becomes zero when $p_{\bm{\lambda}}$ matches the Boltzmann distribution, provided that mode collapse is avoided~\cite{Wu_2019}. 

The gradient updates are implemented using the Adam optimizer~\cite{kingma2014adam}. Furthermore, the computational complexity of VCA for one gradient descent step is $\mathcal{O}(N_s \times N \times d_h^2)$ for 1D RNNs and 2D RNNs (both vanilla and tensorized versions) and $\mathcal{O}(N_s \times N \log(N) \times d_h^2)$ for dilated RNNs. Consequently, VCA has lower computational cost than VQA, which is implemented using VMC (see Methods Sec.~\ref{app:VMC}).

Finally, we note that in our implementations no training steps are performed at the end of annealing for both VCA and VQA. 

\subsection{Variational Monte Carlo}
\label{app:VMC}

The main goal of Variational Monte Carlo is to approximate the ground state of a Hamiltonian $\hat{H}$ through the iterative optimization of an ansatz wave function $\ket{\Psi_{\bm{\lambda}}}$. The VMC objective function is given by
\begin{align*}
  E \equiv \frac{\braket{\Psi_{\bm{\lambda}}|\hat{H}|\Psi_{\bm{\lambda}}}}{\braket{\Psi_{\bm{\lambda}}|\Psi_{\bm{\lambda}}}}.
  \label{eq:VMCenergy}
\end{align*}
We note that an important class of stoquastic many-body
Hamiltonians has ground states $\ket{\Psi}$ with strictly real and positive amplitudes in the standard product spin basis~\cite{Bravyi:2008:CSL:2011772.2011773}. These ground states can be written down in terms of probability distributions, 
\begin{align}
    \ket{\Psi} = \sum_{\bm{\sigma}} \Psi(\bm{\sigma})\ket{\bm{\sigma}} = \sum_{\bm{\sigma}}\sqrt{P(\bm{\sigma})}\ket{\bm{\sigma}}.
\end{align}
To approximate this family of states, we use an RNN wave function, namely $\Psi_{\bm{\lambda}}(\bm{\sigma})  = \sqrt{p_{\bm{\lambda}}(\bm{\sigma})}$. Extensions to complex-valued RNN wave functions are defined in Ref.~\cite{RNNWF_2020}, and results on their ability to simulate variational quantum annealing of non-stoquastic Hamiltonians~\cite{experimental_nonstoq2020} will be reported elsewhere~\cite{vqa_nonstoq}. These families of RNN states are normalized by construction (i.e., $\braket{\Psi_{\bm{\lambda}}|\Psi_{\bm{\lambda}}} = 1$) and allow for accurate estimates of the energy expectation value. By taking $N_s$ exact samples $\bm{\sigma}^{(i)} \sim p_{\bm{\lambda}}$  ($i=1,\ldots,N_s$), it follows that 
\begin{equation*}
    E \approx \frac{1}{N_s} \sum_{i=1}^{N_s}  E_{\rm loc}(\bm{\sigma}^{(i)}).
\end{equation*}
The local energy is given by 
\begin{equation}
E_{\rm loc}(\bm{\sigma}) = \sum_{\bm{\sigma'}} H_{\bm{\sigma\sigma'}}\frac{\Psi_{\bm{\lambda}}(\bm{\sigma'})}{\Psi_{\bm{\lambda}}(\bm{\sigma})},
\label{eq:local_energy}
\end{equation}
where the  sum over $\bm{\sigma'}$ is tractable when the Hamiltonian $\hat{H}$ is local. Similarly, we can also estimate the energy gradients as
\begin{equation*}
    \partial_{\bm{\lambda}} E = \frac{2}{N_s}  \sum_{i=1}^{N_s} \partial_{\bm{\lambda}} \log \left( \Psi^{}_{\bm{\lambda}}\left(\bm{\sigma}^{(i)}\right) \right) \left ( E_{\rm loc}\left(\bm{\sigma}^{(i)}\right) - E \right ).
\end{equation*}
Here, we can subtract the term $E$ in order to reduce noise in the stochastic estimation of our gradients without introducing a bias~\cite{mohamed2019monte, RNNWF_2020}. In fact, when the ansatz is close to an eigenstate of $\hat{H}$, then $E_{\rm loc}(\bm{\sigma}) \approx E$, which means that the variance of gradients $\text{Var}(\partial_{\lambda_j} E) \approx 0$ for each variational parameter $\lambda_j$. We note that this is similar in spirit to the control variate methods in Monte Carlo and to the baseline methods in reinforcement learning~\cite{mohamed2019monte}.

Similarly to the minimization scheme of the variational free energy in Methods Sec.~\ref{app:VCA}, VMC also exhibits a zero-variance principle, where the energy variance per spin
\begin{equation}
    \sigma^2 \equiv \frac{\text{var}(\{E_{\rm loc}(\bm{\sigma})\})}{N},
    \label{eq:energyvariance}
\end{equation}
becomes zero when $\ket{\Psi_{\rm \bm{\lambda}}}$ matches an excited state of $\hat{H}$, which thanks to the minimization of the variational energy $E$ is likely to be the ground state $\ket{\Psi_{\rm G}}$. 

The gradients $\partial_{\bm{\lambda}} \log \left( \Psi^{}_{\bm{\lambda}}\left(\bm{\sigma^{}}\right) \right)$ are numerically computed using automatic  differentiation~\cite{zhang2019automatic}. We use the Adam optimizer to perform gradient descent updates, with a learning rate $\eta$, to optimize the variational parameters $\bm{\lambda}$ of the RNN wave function. We note that in the presence of $\mathcal{O}(N)$ non-diagonal elements in a Hamiltonian $\hat{H}$, the local energies $E_{\rm loc}(\bm{\sigma})$ have $\mathcal{O}(N)$ terms (see Eq.~\eqref{eq:local_energy}). Thus, the computational complexity of one gradient descent step is $\mathcal{O}(N_s \times N^2 \times d_h^2)$ for 1D RNNs and 2D RNNs (both vanilla and tensorized versions).


\subsection{Simulated Quantum Annealing and Simulated Annealing}
\label{app:PIQMC}
Simulated Quantum Annealing is a standard quantum-inspired classical technique that has traditionally been used to benchmark the behavior of quantum annealers~\cite{boixo2014evidence}. It is usually implemented via the path-integral Monte Carlo method~\cite{Santoro_2002}, a QMC method that simulates equilibrium properties of quantum systems at finite temperature. To illustrate this method, consider a $D$-dimensional time-dependent quantum Hamiltonian 
\begin{equation*}
    \hat{H}(t) = - \sum_{i,j} J_{ij} \hat{\sigma}_i^z \hat{\sigma}_j^z - \Gamma(t) \sum_{i=1}^{N} \hat{\sigma}_i^{x},
\end{equation*}
where $\Gamma(t) = \Gamma_0 (1-t)$ controls the strength of the quantum annealing dynamics at a time $t\in[0,1]$. By applying the Suzuki-Trotter formula to the partition function of the quantum system, 
\begin{equation}
    Z = \text{Tr}\exp\{-\beta \hat{H}(t)\} 
    \label{eq:partitionfun},
\end{equation}
with the inverse temperature $\beta = \frac{1}{T}$, we can map the $D$-dimensional quantum Hamiltonian onto a $(D+1)$ classical system consisting of $P$ coupled replicas (Trotter slices) of the original system
\begin{equation}
    H_{D+1}(t) = -\sum_{k=1}^{P} \left ( \sum_{i,j} J_{ij} \sigma_{i}^k \sigma_{j}^k + J_{\perp}(t) \sum_{i=1}^{N} \sigma_{i}^k \sigma_{i}^{k+1} \right ),
    \label{eq:trotterized_hamiltonian}
\end{equation}
where $\sigma_{i}^k$ is the classical spin at site $i$ and replica $k$.  The term $J_{\perp}(t)$ corresponds to uniform coupling between $\sigma_{i}^k$ and $\sigma_{i}^{k+1}$ for each site $i$, such that
\begin{equation*}
    J_{\perp}(t) = - \frac{PT}{2} \ln \left ( \tanh \left(\frac{\Gamma(t)}{PT}\right)  \right ).
\end{equation*}
We note that periodic boundary conditions $\sigma^{P+1}\equiv\sigma^1$ arise because of the trace in Eq.~\eqref{eq:partitionfun}. 

Interestingly, we can approximate $Z$ with an effective partition function $Z_p$ at temperature $PT$ given by~\cite{Santoro2002_2}:
\begin{equation*}
    Z_p \propto \text{Tr}\exp\left\{-\frac{H_{\rm D+1}(t)}{PT} \right\},
\end{equation*}  
which can now be simulated with a standard Metropolis-Hastings Monte Carlo algorithm. A key element to this algorithm is the energy difference induced by a single spin flip at site $\sigma^k_i$, which is equal to 
\begin{align*}
    \Delta_i E_{\rm local} = 2 \sum_{j} J_{ij} \sigma_i^k \sigma_j^k + 2J_{\perp}(t)\left(\sigma_i^{k-1} \sigma_i^{k} + \sigma_i^k \sigma_i^{k+1}\right).
\end{align*}
Here, the second term encodes the quantum dynamics. In our simulations we consider single spin flip (local) moves applied to all sites in all slices. We can also perform a global move~\cite{Santoro2002_2}, which means flipping a spin at location $i$ in every slice $k$. Clearly this has no impact on the term dependent on $J_\perp$, because it contains only terms quadratic in the flipped spin, so that
\begin{align*} 
    \Delta_i E_{\rm global} = 2 \sum_{k=1}^P\sum_{j} J_{ij} \sigma_i^k \sigma_j^k.
\end{align*}
In summary, a single Monte Carlo step (MCS) consists of first performing a single local move on all sites in each $k$-th slice and on all slices, followed by a global move for all sites. For the SK model and the WPE model studied in this paper, we use $P = 100$, whereas for the EA model we use $P = 20$ similarly to Ref.~\cite{Santoro_2002}. 
Before starting the quantum annealing schedule, we first thermalize the system by performing SA~\cite{Santoro2002_2} from a temperature $T_0 = 3$ to a final temperature $1/P$ (so that $PT = 1$). This is done in $60$ steps, where at each temperature we perform $100$ Metropolis moves on each site. 
We then perform SQA using a linear schedule that decreases the field from $\Gamma_0$ to a final value close to zero $\Gamma(t=1)=10^{-8}$, where five local and global moves are performed for each value of the magnetic field $\Gamma (t)$, so that it is consistent with the choice of $N_{\rm train} = 5$ for VCA (see Sec.~\ref{sec:VCA} and~\ref{sec:random_ising_chains}). Thus, the number of MCS is equal to five times the number of annealing steps.

For the standalone SA, we decrease the temperature from $T_0$ to $T(t=1) = 10^{-8}$. Here, a single MCS consists of a Monte Carlo sweep, i.e., attempting a spin-flip for all sites. For each thermal annealing step, we perform five MCS, and hence similar to SQA, the number of MCS is equal to fives times the number of annealing steps. Furthermore, we do a warm-up step for SA, by performing $N_{\rm warmup}$ MCS to equilibrate the Markov Chain at the initial temperature $T_0$ and to provide a consistent choice with VCA (see Sec.~\ref{sec:VCA}).


% ...


\section*{Acknowledgments}
We acknowledge Jack Raymond for suggesting to use the Wishart Planted Ensemble as a benchmark for our variational annealing setup. We also thank Christopher Roth, Cunlu Zhou, Martin Ganahl and Giuseppe Santoro for fruitful discussions. We are also grateful to Lauren Hayward for providing her plotting code to produce our figures using Matplotlib library. Our RNN implementation is based on Tensorflow and NumPy. We acknowledge support from the Natural Sciences and Engineering Research Council (NSERC), a Canada Research Chair, the Shared Hierarchical Academic Research Computing Network (SHARCNET), Compute Canada, Google Quantum Research Award, and the Canadian Institute for Advanced Research (CIFAR) AI chair program. Resources used in preparing this research were provided, in part, by the Province of Ontario, the Government of Canada through CIFAR, and companies sponsoring the Vector Institute \url{www.vectorinstitute.ai/#partners}. Research at Perimeter Institute is supported in part by the Government of Canada through the Department of Innovation, Science and Economic Development Canada and by the Province of Ontario through the Ministry of Economic Development, Job Creation and Trade.

\appendix 
\section{Numerical proof of principle of adiabaticity}
\label{app:adiabaticity_proof}

\begin{figure}
    \centering
    \includegraphics[width =\linewidth]{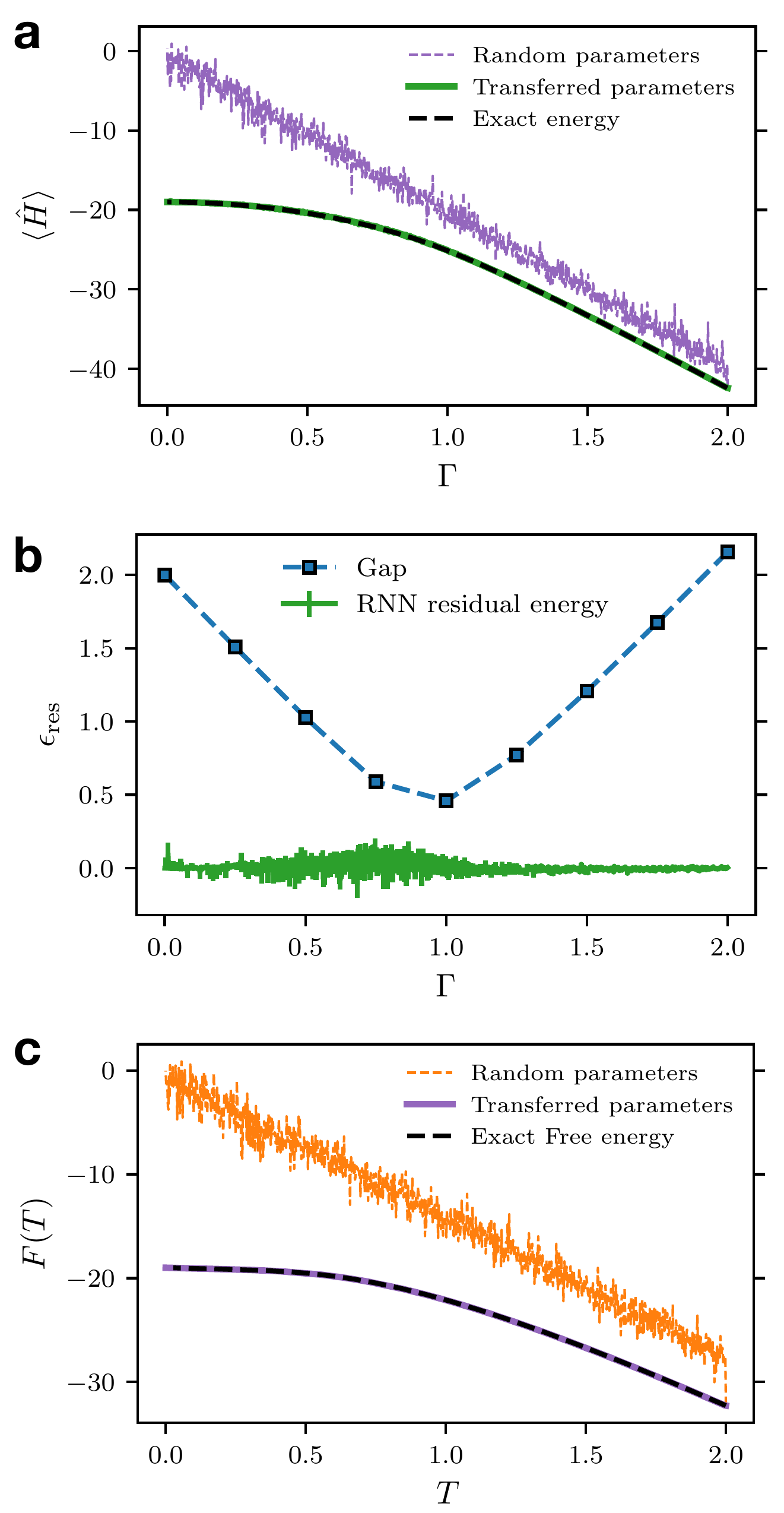}
    \caption{Numerical evidence of adiabaticity on the uniform Ising chain with $N=20$ spins for VQA in panels (a) and (b) and VCA in panel (c). (a) Variational energy of RNN wave function against the transverse magnetic field $\Gamma$, with $\bm{\lambda}$ initialized using the parameters optimized in the previous annealing step (transferred parameters, green curve) and with random parameter reinitialization (random parameters, purple curve). These strategies are compared with the exact energy obtained from exact diagonalization (dashed black line). (b) Residual energy of the RNN wave function vs the transverse field $\Gamma$. Throughout annealing with VQA, the residual energy is always much smaller than the gap within error bars. (c) Variational free energy vs temperature $T$ for a VCA run with $\bm{\lambda}$ initialized using the parameters optimized in the previous annealing step  (transferred parameters, purple line) and with random reinitialization (random parameters, orange curve).}
    \label{fig:adiabaticitycurve}
\end{figure}

As demonstrated in Sec.~\ref{sec:results}, we have shown that both VQA and VCA are effective at finding the classical ground state of disordered spin chains. Here, we further illustrate the adiabaticity of both VQA and VCA. First, we perform VQA on the uniform ferromagnetic Ising chain (i.e., $J_{i,i+1} = 1$) with $N = 20$ spins and open boundary conditions with an initial transverse field $\Gamma_0 = 2$. Here, we use a tensorized RNN wave function with weight sharing across sites of the chain. We also choose $N_{\rm annealing} = 1024$. In Fig.~\ref{fig:adiabaticitycurve}(a), we show that the variational energy tracks the exact ground energy throughout the annealing process with high accuracy. We also observe that optimizing an RNN wave function from scratch, i.e., randomly reinitializing the parameters of the model at each new value of the transverse magnetic field is not optimal. This observation underlines the importance of transferring the parameters of our wave function ansatz after each annealing step. Furthermore, in Fig.~\ref{fig:adiabaticitycurve}(b) we illustrate  that the RNN wave function's residual energy is much lower compared to the gap throughout the annealing process, which shows that VQA remains adiabatic for a large number of annealing steps. 

Similarly, in Fig.~\ref{fig:adiabaticitycurve}(c) we perform VCA with an initial temperature $T_0 = 2$ on the same model, the same system size, the same ansatz, and the same number of annealing steps. We see an excellent agreement between the RNN wave function free energy and the exact free energy, highlighting once again the adiabaticity of our emulation of classical annealing, as well as the importance of transferring the parameters of our ansatz after each annealing step. Taken all together, the results in Fig.~\ref{fig:adiabaticitycurve} support the notion that VQA and VCA evolutions can be adiabatic.

\begin{figure}
    \centering
    \includegraphics[width =\linewidth]{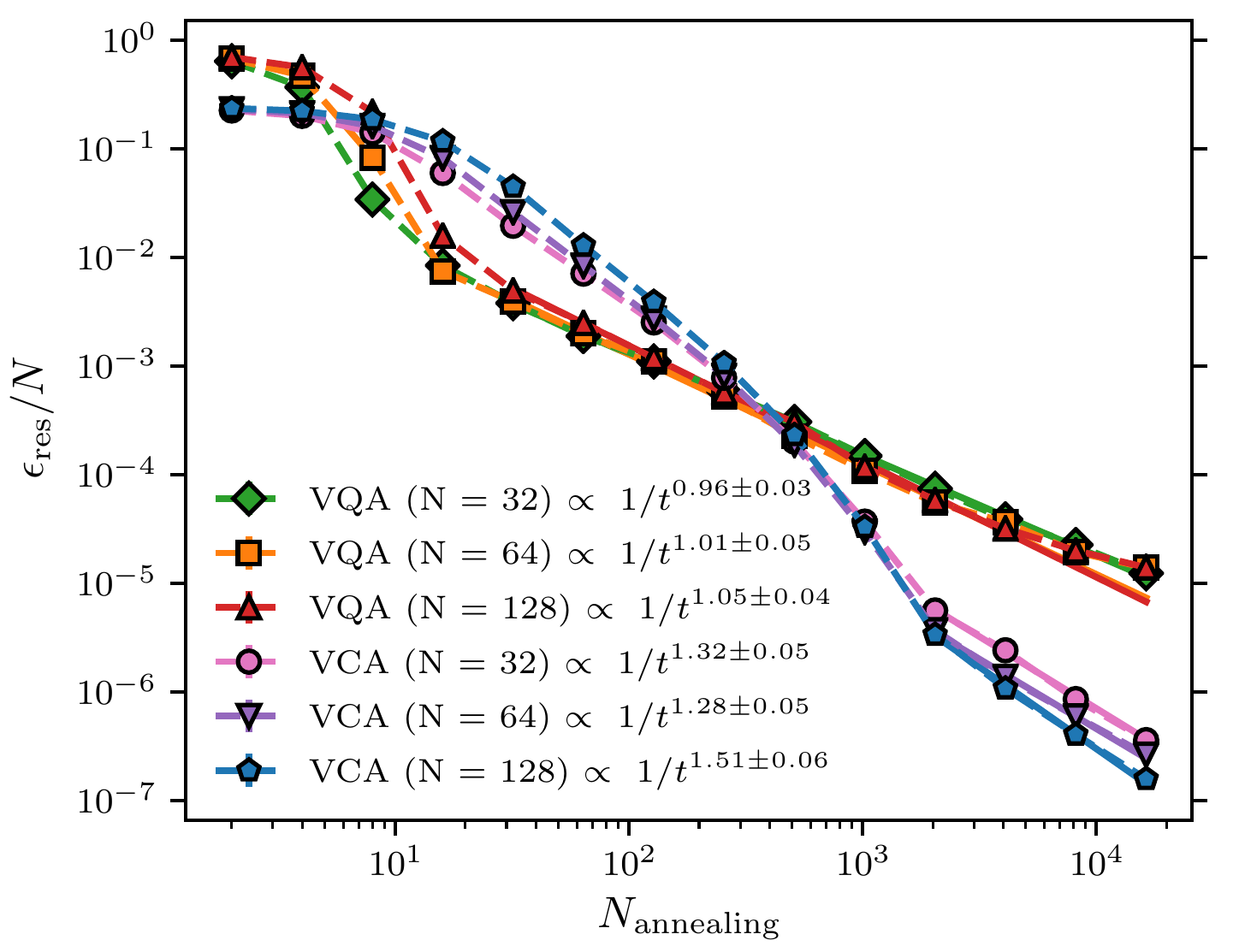} 
    \caption{Variational annealing on random Ising chains, where we represent the residual energy per site $\epsilon_{\rm res}/N$ vs $N_{\rm annealing}$ for both VQA and VCA. The system sizes are $N = 32,64,128$ and we use random discrete couplings $J_{i,i+1} \in \{-1,1\}$.
    } 
    \label{fig:1DscalingB}
\end{figure}

In Fig.~\ref{fig:1DscalingB} we report the residual energies per site against the number of annealing steps $N_{\rm annealing}$. Here, we consider $J_{i,i+1}$ uniformly sampled from the discrete set $\{-1,+1\}$,  where the ground state configuration is disordered and the ground state energy is given by $E_G = -\sum_{i=1}^{N-1} |J_{i,i+1}| = -(N-1)$.  The decay exponents for VCA are in the interval $1.3-1.6$ and the VQA exponent are approximately $1$. These exponents also suggest an asymptotic speed-up compared to SA and coherent quantum annealing, where the residual energies follow a logarithmic law~\cite{Suzuki_2009, dziarmaga20006, caneva2007, PhysRevB.93.224431}. The latter confirms the robustness of the observations in Fig.~\ref{fig:1Dscaling}.

\section{The variational adiabatic theorem}
\label{app:derivation_annealingtime}
In this section, we derive a sufficient condition for the number of gradient descent steps needed to maintain the variational ansatz close to the instantaneous ground state throughout the VQA simulation. First, consider a variational wave function $\ket{\Psi_{\bm \lambda}}$ and the following the time-dependent Hamiltonian:
\begin{equation*}
    \hat{H}(t) = \hat{H}_{\text{target}} + f(t) \hat{H}_D,
\end{equation*}
The goal is to find the ground state of the target Hamiltonian $\hat{H}_{\text{target}}$ by introducing quantum fluctuations through a driving Hamiltonian $\hat{H}_D$, where $\hat{H}_D \gg \hat{H}_{\text{target}}$. Here $f(t)$ is a decreasing schedule function such that $f(0) = 1$, $f(1) = 0$ and $t \in [0,1]$.

Let $E(\bm{\lambda},t) = \bra{\Psi_{\bm{\lambda}}} \hat{H}(t) \ket{\Psi_{\bm{\lambda}}}$, and $E_G(t), E_E(t)$  the instantaneous ground/excited state energy of the Hamiltonian $\hat{H}(t)$, respectively. The instantaneous energy gap is defined as $g(t) \equiv E_E(t)- E_G(t)$.

To simplify our discussion, we consider the case of a target Hamiltonian that has a non-degenerate ground state. Here, we decompose the variational wave function as:
\begin{equation}
    \ket{\Psi_{\bm \lambda}} = (1-a(t))^{\frac12} \ket{\Psi_{G}(t)} + a(t)^{\frac12} \ket{\Psi_{\perp}(t)},
\end{equation}
where $\ket{\Psi_{G}(t)}$ is the instantaneous ground state and $\ket{\Psi_{\perp}(t)}$ is a superposition of all the instantaneous excited states. From this decomposition, one can show that~\cite{sorella2013sissa}:
\begin{equation}
    a(t) \leq \frac{E(\bm{\lambda}, t) - E_G(t)}{g(t)}.
    \label{eq:restogap_ratio}
\end{equation}
As a consequence, in order to satisfy adiabaticity, i.e., $|\left \langle \Psi_{\perp}(t)|\Psi_{\bm{\lambda}} \right \rangle|^2 \ll 1$ for all times $t$, then one should have $a(t) < \epsilon \ll 1$ where $\epsilon$ is a small upper bound on the overlap between the variational wave function and the excited states. This means that the success probability $P_{\rm success}$ of obtaining the ground state at $t=1$ is bounded from below by $1-\epsilon$. From Eq.~\eqref{eq:restogap_ratio}, to satisfy $a(t) < \epsilon$, it is sufficient to have:
\begin{equation}
    \epsilon_{\rm res}(\bm{\lambda}, t) \equiv E(\bm{\lambda}, t) - E_G(t) < \epsilon g(t).
    \label{eq:weakcondition}
\end{equation}
To satisfy the latter condition, we require a slightly stronger condition as follows:
\begin{equation}
      \epsilon_{\rm res}(\bm{\lambda}, t)  < \frac{\epsilon g(t)}{2}.
     \label{eq:strong_condition}
\end{equation}
In our derivation of a sufficient condition on the number of gradient descent steps to satisfy the previous requirement, we use the following set of assumptions:
\begin{itemize}
    \item \textbf{(A1)} $|\partial^k_t E_G(t)|, |\partial^k_t g(t)|, |\partial^k_t f(t)| \leq \mathcal{O}(\text{poly}(N))$, for all $0 \leq t \leq 1$ and for $k \in \{1,2\}$.
    \item \textbf{(A2)} $| \langle \Psi_{\bm{\lambda}} | \hat{H}_D | \Psi_{\bm{\lambda}} \rangle | \leq \mathcal{O}(\text{poly}(N))$ for all possible parameters $\bm{\lambda}$ of the variational wave function.
    \item \textbf{(A3)} No anti-crossing during annealing, i.e., $g(t) \neq 0$, for all $0 \leq t \leq 1$.
    \item \textbf{(A4)} The gradients $\partial_{\bm{\lambda}} E({\bm{\lambda}}, t)$ can be calculated exactly, are $L(t)$-Lipschitz with respect to $\bm{\lambda}$ and $L(t) \leq \mathcal{O}(\text{poly}(N))$ for all $0 \leq t \leq 1$.
    \item \textbf{(A5)} Local convexity, i.e., close to convergence when $\epsilon_{\rm res}(\bm{\lambda}, t) < \epsilon g(t)$, the energy landscape of $E({\bm{\lambda}}, t)$ is convex with respect to $\bm{\lambda}$, for all $0 < t \leq 1$. 
    
    Note that this assumption is $\epsilon$-dependent.
    \item \textbf{(A6)} The parameters vector $\bm{\lambda}$ is bounded by a polynomial in $N$. i.e., $||\bm{\lambda}|| \leq \mathcal{O}(\text{poly}(N))$, where we define ``$||.||$'' as the euclidean $L_2$ norm.
    \item \textbf{(A7)} The variational wave function $|\Psi_{\bm{\lambda}} \rangle$ is expressive enough, i.e.,
    \begin{equation*}
    \displaystyle\min_{\bm{\lambda}} \epsilon_{\rm res}(\bm{\lambda}, t) < \frac{ \epsilon g(t)}{4}, \quad \forall t \in [0,1].
\end{equation*}
    Note that this assumption is also $\epsilon$-dependent.
    \item \textbf{(A8)} At $t = 0$, the energy landscape of $E(\bm{\lambda}, t = 0)$ is globally convex with respect to $\bm{\lambda}$. 
    
\end{itemize}

\textbf{Theorem} Given the assumptions \textbf{(A1)} to \textbf{(A8)}, a sufficient (but not necessary) number of gradient descent steps $N_{\rm steps}$ to satisfy the condition~\eqref{eq:strong_condition} during the VQA protocol, is bounded as:
\begin{equation*}
     \mathcal{O} \left(\frac{\text{poly}(N)}{\epsilon \displaystyle\min_{\{t_n\}}(g(t_n))} \right) \leq N_{\text{steps}} \leq \mathcal{O} \left(\frac{\text{poly}(N)}{\epsilon^2 \displaystyle\min_{\{t_n\}}(g(t_n))^2} \right),
\end{equation*}
where $(t_1,t_2, t_3, \ldots)$ is an increasing finite sequence of time steps, satisfying $t_1 = 0$ and $t_{n+1} = t_n + \delta t_n$, where
\begin{equation*}
    \delta t_n = \mathcal{O}\left ( \frac{\epsilon g(t_n)}{\text{poly}(N)} \right).
\end{equation*}

\textbf{Proof:}
In order to satisfy the condition Eq.~\eqref{eq:strong_condition} during the VQA protocol, we follow these steps:
\begin{itemize}
    \item Step 1 (warm-up step): we prepare our variational wave function at the ground state at $t = 0$ such that Eq.~\eqref{eq:strong_condition} is verified at time $t = 0$.
    \item Step 2 (annealing step): we change time $t$ by an infinitesimal amount $\delta t$, so that the condition~\eqref{eq:weakcondition} is verified at time $t+\delta t$.
    \item Step 3 (training step): we tune the parameters of the variational wave function, using gradient descent, so that the condition~\eqref{eq:strong_condition} is satisfied at time $t+\delta t$.
    \item Step 4: we loop over steps 2 and 3 until we arrive at $t = 1$, where we expect to obtain the ground state energy of the target Hamiltonian.
\end{itemize}
Let us first start with step 2 assuming that step 1 is verified. In order to satisfy the requirement of this step at time $t$, then $\delta t$ has to be chosen small enough so that
\begin{equation}
    \epsilon_{\rm res}(\bm{\lambda}_t, t+\delta t) < \epsilon g(t+\delta t)
    \label{eq:annealingcondition}
\end{equation}
is verified given that the condition~\eqref{eq:strong_condition} is satisfied at time $t$. Here, $\bm{\lambda}_t$ are the parameters of the variational wave function that satisfies the condition~\eqref{eq:strong_condition} at time $t$. To get a sense of how small $\delta t$ should be, we do a Taylor expansion, while fixing the parameters $\bm{\lambda}_t$, to get:
\begin{align*}
    & \epsilon_{\rm res}(\bm{\lambda}_t, t+\delta t) \\
    &= \epsilon_{\rm res}(\bm{\lambda}_t, t) + \partial_t \epsilon_{\rm res}(\bm{\lambda}_t, t)\delta t + \mathcal{O}((\delta t)^2),\\
    &< \frac{\epsilon g(t)}{2} + \partial_t \epsilon_{\rm res}(\bm{\lambda}_t, t)\delta t + \mathcal{O}((\delta t)^2),
\end{align*}
where we used the condition~\eqref{eq:strong_condition} to go from the second line to the third line. Here, $\partial_t \epsilon_{\rm res}(\bm{\lambda}_t, t) = \partial_t f(t)  \langle \hat{H}_D \rangle - \partial_t E_G(t)$. To satisfy the condition~\eqref{eq:weakcondition} at time $t+\delta t$, it is enough to have the right hand side of the previous inequality to be much smaller than the gap at $t+\delta t$, i.e.,
\begin{equation*}
    \frac{\epsilon g(t)}{2} + \partial_t \epsilon_{\rm res}(\bm{\lambda}_t, t)\delta t + \mathcal{O}((\delta t)^2) < \epsilon g(t+\delta t).
\end{equation*}
By Taylor expanding the gap, we get:
\begin{align*}
    \partial_t \epsilon_{\rm res}(\bm{\lambda}_t, t)\delta t + \mathcal{O}((\delta t)^2) < \frac{\epsilon g(t)}{2} + \epsilon \partial_t g(t) \delta t + \mathcal{O}((\delta t)^2),
\end{align*}
hence, it is enough to satisfy the following condition:
\begin{equation}
    (\partial_t \epsilon_{\rm res}(\bm{\lambda}_t, t) - \epsilon \partial_t g(t))\delta t + \mathcal{O}((\delta t)^2) < \frac{\epsilon g(t)}{2}.
    \label{eq:eq1}
\end{equation}
Using the Taylor-Laplace formula, one can express the Taylor remainder term $\mathcal{O}((\delta t)^2)$ as follows:
\begin{equation*}
    \mathcal{O}((\delta t)^2) = \int_{t}^{t+\delta t}  (\tau-t) A(\tau) {\rm d}\tau,
\end{equation*}
where $A(\tau) = \partial^2_\tau \epsilon_{\rm res}(\bm{\lambda}_t, \tau) - \epsilon \partial^2_\tau g(\tau) = \partial^2_\tau f(\tau)  \langle \hat{H}_D \rangle - \partial^2_\tau E_G(\tau) - \epsilon \partial^2_\tau g(\tau)$ and $\tau$ is between $t$ and $t+\delta t$. The last expression can be bounded as follows:
\begin{equation*}
    \mathcal{O}((\delta t)^2) \leq \int_{t}^{t+\delta t} (\tau-t)  |A(\tau)| {\rm d}\tau \leq \frac{(\delta t)^2}{2} {\rm sup}(|A|).
\end{equation*}
where ``$\rm{sup}(|A|)$'' is the supremum of $|A|$ over the interval $[0,1]$. Given assumptions \textbf{(A1)} and \textbf{(A2)}, then $\rm{sup}(|A|)$ is bounded from above by a polynomial in $N$, hence:
\begin{equation*}
    \mathcal{O}((\delta t)^2) \leq \mathcal{O}(\text{poly}(N)) (\delta t)^2 \leq \mathcal{O}(\text{poly}(N)) \delta t,
\end{equation*}
where the last inequality holds since $\delta t \leq 1$ as $t \in [0,1]$, while we note that it is not necessarily tight. 
Furthermore, since $(\partial_t \epsilon_{\rm res}(\bm{\lambda}_t, t)   - \epsilon \partial_t g(t))$ is also bounded from above by a polynomial in $N$ (according to assumptions \textbf{(A1)} and \textbf{(A2)}), then in order to satisfy Eq.~\eqref{eq:eq1}, it is sufficient to require the following condition:
\begin{equation*}
    \mathcal{O}(\text{poly}(N)) \delta t < \frac{\epsilon g(t)}{2}.
\end{equation*}
Thus, it is sufficient to take:
\begin{equation}
    \delta t = \mathcal{O}\left ( \frac{\epsilon g(t)}{\text{poly}(N)} \right).
    \label{eq:step_condition}
\end{equation}
By taking account of assumption (\textbf{A3}), $\delta t$ can be taken non-zero for all time steps $t$. As a consequence, assuming the condition~\eqref{eq:step_condition} is verified for a non-zero $\delta t$ and a suitable $\mathcal{O}(1)$ prefactor, then the condition~\eqref{eq:annealingcondition} is also verified. 

We can now move to step 3. Here, we apply a number of gradient descent steps $N_{\rm train}(t)$ to find a new set of parameters $\bm{\lambda}_{t+\delta t}$ such that:
\begin{equation}
    \epsilon_{\rm res} (\bm{\lambda}_{t+\delta t}, t+\delta t) = E(\bm{\lambda}_{t+\delta t}, t+\delta t) - E_G(t+\delta t) < \frac{\epsilon g(t+\delta t)}{2},
    \label{eq:trainingcondition}
\end{equation}
To estimate the scaling of the number of gradient descent steps $N_{\rm train}(t)$ needed to satisfy~\eqref{eq:trainingcondition}, we make use of assumptions \textbf{(A4)} and \textbf{(A5)}. The assumption \textbf{(A5)} is reasonable providing that the variational energy $E(\bm{\lambda}_t, t+\delta t)$ is very close to the ground state energy $E_G(t+\delta t)$, as given by Eq.~\eqref{eq:annealingcondition}. Using the above assumptions and assuming that the learning rate $\eta(t) = 1/L(t)$, we can use a well-known result in convex optimization~\cite{Nesterov2018}(see Sec.~2.1.5), which states the following inequality:
\begin{align*}
    E(\bm{\tilde{\lambda}}_{t}, t+\delta t) - \displaystyle\min_{\bm{\lambda}} E(\bm{\lambda}, t+\delta t) \leq \frac{2 L(t) ||\bm{\lambda}_t - \bm{\lambda}^{*}_{t+\delta t}||^2}{N_{\rm train}(t)+4}.
\end{align*}
Here, $\bm{\tilde{\lambda}}_{t}$ are the new variational parameters obtained after applying $N_{\rm train}(t+\delta t)$ gradient descent steps starting from $\bm{\lambda}_t$. Furthermore, $\bm{\lambda}^{*}_{t+\delta t}$ are the optimal parameters such that:
\begin{equation*}
    E(\bm{\lambda}^{*}_{t+\delta t},t+\delta t) = \min_{\bm{\lambda}}  E(\bm{\lambda}, t+\delta t).
\end{equation*}
Since the Lipschitz constant $L(t) \leq \mathcal{O}(\text{poly}(N))$ (assumption \textbf{(A4)}) and $ ||\bm{\lambda}_t - \bm{\lambda}^{*}_{t+\delta t}||^2 \leq \mathcal{O}(\text{poly}(N))$ (assumption \textbf{(A6)}), one can take
\begin{equation}
N_{\rm train}(t+\delta t) = \mathcal{O}\left( \frac{\text{poly}(N)}{\epsilon g(t+\delta t)} \right),
\label{eq:numgradsteps}
\end{equation}
with a suitable $\mathcal{O}(1)$ prefactor, so that:
\begin{equation*}
    E(\bm{\tilde{\lambda}}_{t}, t+\delta t) - \displaystyle\min_{\bm{\lambda}} E(\bm{\lambda}, t+\delta t) < \frac{ \epsilon g(t+\delta t)}{4}.
\end{equation*}
Moreover, by assuming that the variational wave function is expressive enough (assumption \textbf{(A7)}), i.e.,
\begin{equation*}
    \displaystyle\min_{\bm{\lambda}} E(\bm{\lambda}, t+\delta t) - E_G(t+\delta t) < \frac{ \epsilon g(t+\delta t)}{4},
\end{equation*}
we can then deduce, by taking $\bm{\lambda}_{t+\delta t} \equiv \bm{\tilde{\lambda}}_{t}$ and summing the two previous inequalities, that:
\begin{equation*}
    E(\bm{\lambda}_{t+\delta t}, t+\delta t) - E_G(t+\delta t)  < \frac{ \epsilon g(t+\delta t)}{2}.
\end{equation*}

Let us recall that in step 1, we have to initially prepare the variational ansatz to satisfy condition~\eqref{eq:strong_condition} at $t=0$. In fact, we can take advantage of the assumption \textbf{(A4)}, where the gradients are $L(0)$-Lipschitz with $L(0) \leq \mathcal{O}(\text{poly}(N))$. We can also use the convexity assumption \textbf{(A8)}, and we can show that a sufficient number of gradient descent steps to satisfy condition~\eqref{eq:strong_condition} at $t = 0$ is estimated as:
\begin{equation*}
    N_{\rm warmup} \equiv N_{\rm train}(0) = \mathcal{O}\left( \frac{\text{poly}(N)}{\epsilon g(0)} \right).
\end{equation*}
The latter can be obtained in a similar way as in Eq.~\eqref{eq:numgradsteps}. 

In conclusion, the total number of gradient steps $N_{\text{steps}}$ to evolve the Hamiltonian $\hat{H}(0)$ to the target Hamiltonian $\hat{H}(1)$, while verifying the condition~\eqref{eq:strong_condition} is given by: 
\[
N_{\rm steps} = \sum_{n=1}^{N_{\rm annealing}+1} N_{\rm train}(t_n),
\]
where each $N_{\rm train}(t_n)$ satisfies the requirement~\eqref{eq:numgradsteps}. The annealing times $\{t_n\}_{n=1}^{N_{\rm annealing}+1}$ are defined such that $t_1 \equiv 0$ and $t_{n+1} \equiv t_n + \delta t_n$. Here, $\delta t_n$ satisfies
\begin{equation}
    \delta t_n =
		   \mathcal{O}\left ( \frac{\epsilon g(t_n)}{\text{poly}(N)} \right).
	   \label{eq:deltat}
\end{equation}
We also consider $N_{\rm annealing}$ the smallest integer such that $t_{N_{\rm annealing}} + \delta t_{N_{\rm annealing}} \geq 1$, in this case, we define $t_{N_{\rm annealing}+1} \equiv 1$, indicating the end of annealing. Thus, $N_{\rm annealing}$ is the total number of annealing steps. Taking this definition into account, then one can show that
\begin{equation*}
    N_{\rm annealing} \leq  \frac{1}{\displaystyle\min_{\{t_n\}} (\delta t_n)} + 1.
\end{equation*}
Using Eqs.~\eqref{eq:step_condition} and~\eqref{eq:numgradsteps} and the previous inequality,
$N_{\rm steps}$ can be bounded from above as:
\begin{align*}
    N_{\text{steps}} & \leq  \left( N_{\rm annealing} + 1 \right) \displaystyle\max_{\{t_n\}}\left(N_{\rm train}(t_n)\right) \\
    & \leq  \left(\frac{1}{\displaystyle\min_{\{t_n\}}(\delta t_n)}+2 \right) \displaystyle\max_{\{t_n\}}\left(N_{\rm train}(t_n)\right) \\ & \leq \mathcal{O}\left(\frac{\text{poly}(N)}{\epsilon^2 \displaystyle\min_{\{t_n\}}(g(t_n))^2}\right),
    \label{eq:upperbound_totalnumgradsteps}
\end{align*}
where the transition from line 2 to line 3 is valid for a sufficiently small $\epsilon$ and $\min_{\{t_n\}}(g(t_n))$. Furthermore, $N_{\text{steps}}$ can also be bounded from below as:
\begin{equation}
    N_{\text{steps}} \geq \displaystyle\max_{\{t_n\}}(N_{\rm train}(t_n)) =  \mathcal{O}\left(\frac{\text{poly}(N)}{\epsilon \displaystyle\min_{\{t_n\}}(g(t_{n}))}\right).
    \label{eq:lowerbound_totalnumgradsteps}
\end{equation}
Note that the minimum in the previous two bounds are taken over all the annealing times $t_n$ where $1 \leq n \leq N_{\rm annealing} + 1$.

In this derivation of the bound on $N_{\rm steps}$, we have assumed that the ground state of $\hat{H}_{\rm target}$ is non-degenerate, so that the gap does not vanish at the end of annealing (i.e., $t = 1$). In the case of degeneracy of the target ground state, we can define the gap $g(t)$ by considering the lowest energy level that does not lead to the degenerate ground state. 

It is also worth noting that the assumptions of this derivation can be further expanded and improved. In particular, the gradients of $E(\bm{\lambda}, t)$ are computed stochastically (see Methods Sec.~\ref{app:VMC}), as opposed to our assumption \textbf{(A4)} where the gradients are assumed to be known exactly. To account for noisy gradients, it is possible to use convergence bounds of stochastic gradient descent~\cite{schmidt2013minimizing,kingma2014adam} to estimate a bound on the number of gradient descent steps. Second-order optimization methods such as stochastic reconfiguration/natural gradient~\cite{becca2017, Amari1998} can potentially show a significant advantage over first-order optimization methods, in terms of scaling with the minimum gap of the time-dependent Hamiltonian $\hat{H}(t)$. 

\section{Default Hyperparameters}
\label{app:hyperparameters}
In this Appendix, we summarize the architectures and the hyperparameters of the simulations performed in this paper, as shown in Tab.~\ref{tab:hyperparams}. The latter has shown to yield good performance, while we believe that a more advanced study of the hyperparameters can result in optimal results. We also note that in this paper, VQA and VCA were run using a single GPU workstation for each simulation, while SQA and SA were performed on a multi-core CPU.

\begin{table*}[p]
    \centering
    \scriptsize
    \begin{tabular}{|c|c|c|}\hline
       Figures & Parameter & Value \\\hline
        \multirow{7}{*}{Figs.~\ref{fig:1Dscaling} and~\ref{fig:1DscalingB}} & Architecture & Tensorized RNN wave function with no-weight sharing \\
            & Number of memory units & $d_h = 40$ \\
            & Number of samples & $N_s = 50$ \\
            & Initial magnetic field for VQA & $\Gamma_0 = 2$ \\
            & Initial temperature for VCA & $T_0 = 1$ \\
            & Learning rate & $\eta = 5 \times 10^{-4}$ \\
            & Warmup steps & $N_{\rm warmup} = 1000$ \\
            & Number of random instances & $N_{\rm instances} = 25$ \\
       \hline
       
      \multirow{8}{*}{Fig.~\ref{fig:EA_comaprison}} & Architecture & 2D tensorized RNN wave function with no weight-sharing \\
            & Number of memory units & $d_h = 40$ \\
            & Number of samples & $N_s = 25$ \\

            & Initial magnetic field & $\Gamma_0 = 1$ (for SQA, VQA and RVQA) \\
            & Initial temperature & $T_0 = 1$ (for SA, VCA and RVQA) \\
            & Learning rate & $\eta = 10^{-4}$ \\
            & Number of warmup steps & $N_{\rm warmup} = 1000$ for $10\times10$ and $N_{\rm warmup} = 2000$ for $40 \times 40$  \\
            & Number of random instances & $N_{\rm instances} = 25$ \\
       \hline
       
          \multirow{8}{*}{Figs.~\ref{fig:SAvsPIQMCvsRNN}(a) and (d)} & Architecture & Dilated RNN wave function with no weight-sharing \\
            & Number of memory units & $d_h = 40$ \\
            & Number of samples & $N_s = 50$  \\
            & Initial temperature  & $T_0 = 2$ (for SA and VCA) \\
            & Initial magnetic field  & $\Gamma_0 = 2$ (for SQA)  \\
            & Learning rate & $\eta = 10^{-4}$ \\
            & Number of warmup steps & $N_{\rm warmup} = 2000$ \\
            & Number of random instances & $N_{\rm instances} = 25$ \\
       \hline
       
         \multirow{8}{*}{Figs.~\ref{fig:SAvsPIQMCvsRNN}(b), (c), (e) and (f)} & Architecture & Dilated RNN wave function with no weight-sharing \\
            & Number of memory units & $d_h = 20$ \\
            & Number of samples & $N_s = 50$ \\
            & Initial temperature & $T_0 = 1$ (for SA and VCA) \\
            & Initial magnetic field & $\Gamma_0 = 1$ (for SQA) \\
            & Learning rate & $\eta = 10^{-4}$ \\
            & Number of warmup steps & $N_{\rm warmup} = 1000$ \\
            & Number of random instances & $N_{\rm instances} = 25$ \\
       \hline
       \multirow{7}{*}{Fig.~\ref{fig:adiabaticitycurve}} & Architecture & Tensorized RNN wave function with weight sharing \\
            & Number of memory units & $d_h = 20$ \\
            & Number of samples & $N_s = 50$ \\
            & Initial temperature & $T_0 = 2$ \\
            & Initial magnetic field & $\Gamma_0 = 2$ \\
            & Learning rate & $\eta = 10^{-3}$ \\
            & Number of warmup steps & $N_{\rm warmup} = 1000$ \\
       \hline
         \multirow{4}{*}{Figs.~\ref{fig:RNNChoice}(a) and (b)} & Architecture & RNN wave function \\
            & Number of memory units & $d_h = 50$  \\
            & Number of samples & $N_s = 50$ \\
            & Learning rate & $\eta = 10^{-3}$ for Fig.~\ref{fig:RNNChoice}(a) and $\eta = 5 \times 10^{-4}$ for Fig.~\ref{fig:RNNChoice}(b)\\
       \hline
       \multirow{5}{*}{Fig.~\ref{fig:RNNChoice}(c)} & Architecture & RNN wave function with no-weight sharing \\
            & Number of memory units of dilated RNN & $d_h = 20$  \\
            & Number of memory units of tensorized RNN & $d_h = 40$  \\
            & Number of samples & $N_s = 100$ \\
            & Learning rate & $\eta = 10^{-4}$\\
       \hline      

    \end{tabular}
    \caption{Hyperparameters used to obtain the results reported in this paper. Note that the number of samples stands for the batch size used to train the RNN.} 
    \label{tab:hyperparams}
\end{table*}

\section{Benchmarking Recurrent neural network cells}
\label{app:benchmarkingRNNcells}
\begin{figure}
    \centering
     \includegraphics[width =\linewidth]{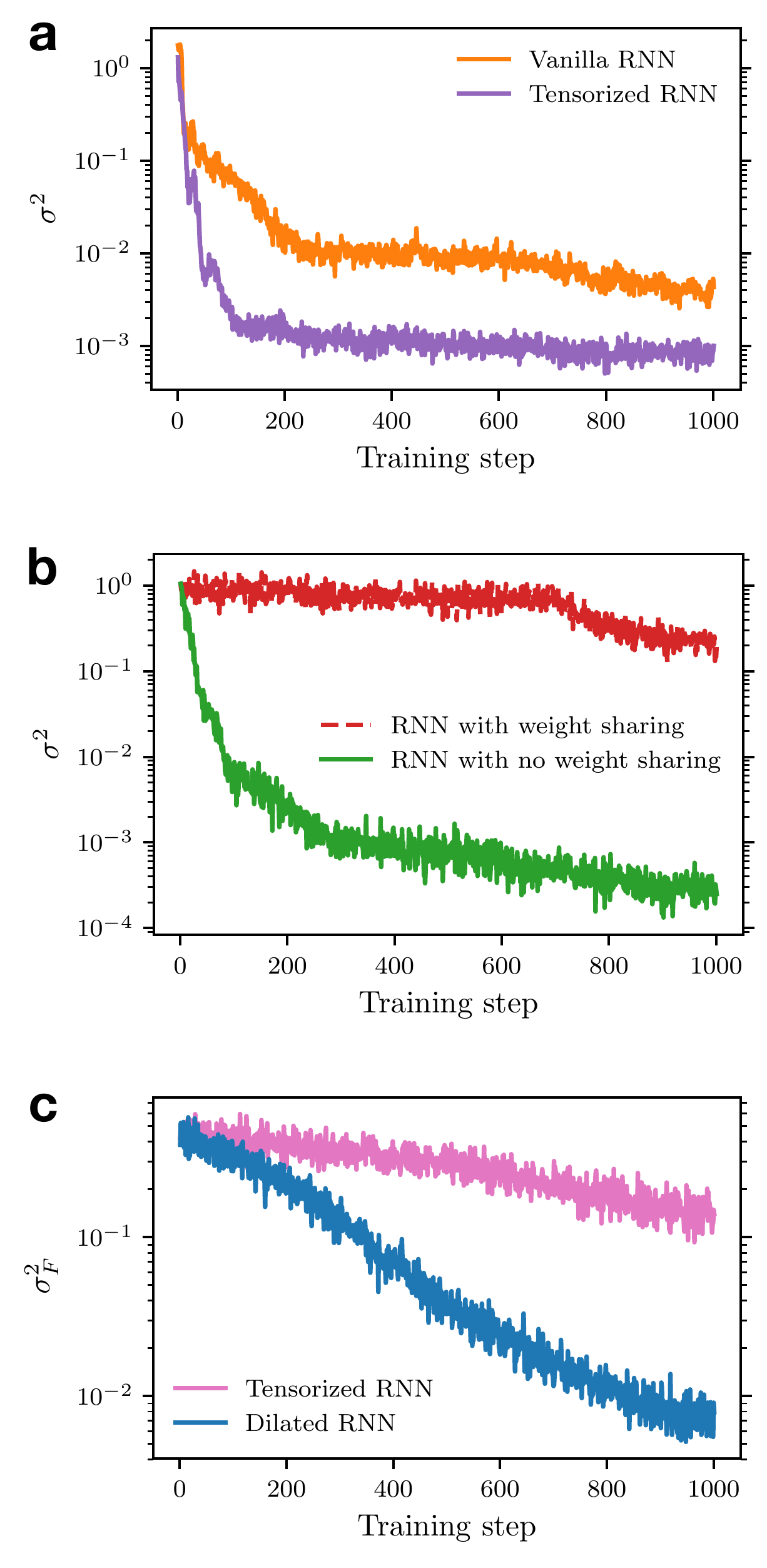}
    \caption{Energy (or Free energy) variance per spin $\sigma^2$ vs the number of training steps. (a) We compare tensorized and vanilla RNN ansatzes both with weight sharing across sites on the  uniform  ferromagnetic Ising  chain at the critical point with $N = 100$ spins. (b) Comparison between a tensorized RNN with and without weight sharing, trained to find the ground state of the random Ising chain with discrete disorder ($J_{i,i+1} = \pm 1$) at criticality with $N = 20$ spins.  (c) Comparison between a tensorized RNN and dilated RNN ansatzes, both with no weight sharing, trained to find the Sherrington-Kirkpatrick model's equilibrium distribution with $N=20$ spins at temperature $T = 1$.}
    \label{fig:RNNChoice}
\end{figure}

 To show the advantage of tensorized RNNs over vanilla RNNs, we benchmark these architectures on the task of finding the ground state of the uniform ferromagnetic Ising chain (i.e., $J_{i,i+1}=1$) with $N = 100$ spins at the critical point (i.e., no annealing is employed). Since the couplings in this model are site-independent, we choose the parameters of the model to be also site-independent. In Fig.~\ref{fig:RNNChoice}(a), we plot the energy variance per site $\sigma^2$ (see Eq.~\eqref{eq:energyvariance}) against the number of gradient descent steps. Here $\sigma^2$ is a good indicator of the quality of the optimized wave function~\cite{Claudius90,Assaraf2003,becca2017}. The results show that the tensorized RNN wave function can achieve both a lower estimate of the energy variance and a faster convergence.

For the disordered systems studied in this paper, we set the weights $T_n, U_n$ and the biases $\bm{b}_n, \bm{c}_n$ (in Eqs.~\eqref{eq:softmax_noweightsharing} and~\eqref{eq:TRNN_recursion}) to be site-dependent. To demonstrate the benefit of using site-dependent over site-independent parameters when dealing with disordered systems, we benchmark both architectures on the task of finding the ground state of the disordered Ising chain with random discrete couplings $J_{i,i+1} = \pm 1$ at the critical point, i.e., with a transverse field $\Gamma = 1$. We show the results in Fig.~\ref{fig:RNNChoice}(b) and find that site-dependent parameters lead to a better performance in terms of the energy variance per spin.

Furthermore, we equally show the advantage of a dilated RNN ansatz compared to a tensorized RNN ansatz. We train both of them for the task of finding the minimum of the free energy of the Sherrington-Kirkpatrick model with $N=20$ spins and at temperature $T = 1$, as explained in Methods Sec.~\ref{app:VCA}. Both RNNs have a comparable number of parameters (66400 parameters for the tensorized RNN and 59240 parameters for the dilated RNN). Interestingly, in Fig.~\ref{fig:RNNChoice}(c), we find that the dilated RNN supersedes the tensorized RNN with almost an order of magnitude difference in term of the free energy variance per spin defined in Eq.~\eqref{eq:freeenergyvariance}. Indeed, this result suggests that the mechanism of skip connections allows dilated RNNs to capture long-term dependencies more efficiently compared to tensorized RNNs.

\clearpage
\bibliography{Biblio}

\end{document}